\PassOptionsToPackage{table}{xcolor}
\documentclass[sigconf]{acmart}
\acmConference[EASE 2025]{The 29th International Conference on Evaluation and Assessment in Software Engineering}{17–20 June, 2025}{Istanbul, Türkiye}

\usepackage{algorithmic}
\usepackage{xcolor}
\usepackage{graphicx}
\usepackage{textcomp}
\usepackage{url}
\usepackage{xspace}
\usepackage{mdframed}
\usepackage{soul,color}
\usepackage{pifont}
\usepackage{float}
\usepackage[inline]{enumitem}
\usepackage{url}

\usepackage{pgfplots}
\usepackage[skip=0pt]{caption}
\usepackage{subcaption}
\usepackage{tikz}

\usepackage{booktabs}
\usepackage{tabularx}
\usepackage{rotating}

\usepackage[most]{tcolorbox}

\newcommand{\ie}{\emph{i.e.,}\xspace}
\newcommand{\eg}{\emph{e.g.,}\xspace}

\newcommand{\etal}{et al.\xspace}

\newlist{questions}{enumerate}{2}
\setlist[questions,1]{label=RQ\arabic*.,ref=RQ\arabic*}
\setlist[questions,2]{label=(\alph*),ref=\thequestionsi(\alph*)}

\newcommand{\RQone}{What is the impact of compilation on the energy efficiency of Python code?}
\newcommand{\RQtwo}{What is the impact of compilation on the performance of Python code?}

\def\BibTeX{{\rm B\kern-.05em{\sc i\kern-.025em b}\kern-.08em
    T\kern-.1667em\lower.7ex\hbox{E}\kern-.125emX}}

\newmdenv[
  linewidth=0pt,
  backgroundcolor=gray!10,
]{resultblock}

\usepackage[most]{tcolorbox}
\definecolor{customblue}{HTML}{45B7D1}
\definecolor{brightBackground}{HTML}{ecf8fa}
\definecolor{plotblue}{HTML}{c5e3ed}

\tcbset{
  colback=plotblue!10,
  colframe=plotblue!80!black,
  left=1mm,
  right=1mm,
  top=1mm,
  bottom=1mm,
  title={},
  boxrule=0.5pt,
  arc=2pt
}

\begin{document}

% How does compilation affect Python performance and energy efficiency? An empirical study
\title{An Empirical Study on the Performance and Energy Usage of Compiled Python Code}

\author{Vincenzo Stoico, Andrei Calin Dragomir, Patricia Lago}
\affiliation{%
 \institution{Vrije Universiteit Amsterdam, The Netherlands}
 \country{v.stoico@vu.nl, a.dragomir@student.vu.nl, p.lago@vu.nl}
}

\begin{abstract}
Python is a popular programming language known for its ease of learning and extensive libraries. However, concerns about performance and energy consumption have led to the development of compilers to enhance Python code efficiency. Despite the proven benefits of existing compilers on the efficiency of Python code, there is limited analysis comparing their performance and energy efficiency, particularly considering code characteristics and factors like CPU frequency and core count. Our study investigates how compilation impacts the performance and energy consumption of Python code, using seven benchmarks compiled with eight different tools: PyPy, Numba, Nuitka, Mypyc, Codon, Cython, Pyston-lite, and the experimental Python 3.13 version, compared to CPython. The benchmarks are single-threaded and executed on an NUC and a server, measuring energy usage, execution time, memory usage, and Last-Level Cache (LLC) miss rates at a fixed frequency and on a single core. The results show that compilation can significantly enhance execution time, energy and memory usage, with Codon, PyPy, and Numba achieving over 90\% speed and energy improvements. Nuitka optimizes memory usage consistently on both testbeds. The impact of compilation on LLC miss rate is not clear since it varies considerably across benchmarks for each compiler. Our study is important for researchers and practitioners focused on improving Python code performance and energy efficiency. We outline future research directions, such as exploring caching effects on energy usage. Our findings help practitioners choose the best compiler based on their efficiency benefits and accessibility.
\end{abstract}

\maketitle

\section{Introduction}
\label{s:introduction}
% Popularity
Regardless of continuous hardware innovations, software ultimately determines how to exploit hardware resources efficiently. As of 2024, it is likely that software running across different devices having different purposes and form factors is written in Python. Python ranks as the third most used language in the Stack Overflow Developer Survey of 2024 with 51\% of the preferences over 60171 respondents \cite{stackoverflow2024}. Due to the extensive support Python provides for third-party libraries, it is adopted in a wide range of domains and purposes. These include artificial intelligence, web development, data analysis, and parallel computing \cite{peng2024less}.

% Inefficiency
The popularity of Python is coupled with recognized limitations concerning its performance and energy efficiency \cite{abdulsalam2014program, georgiou2017analyzing, koedijk2022finding, merelo2016ranking}. Pereira \etal \cite{pereira2021ranking} compares the execution time, memory usage, and energy efficiency of 27 programming languages, including Python, using 10 programming problems implemented in each language. Python always falls in the bottom ranks when the results are sorted by each quality attribute. The results also show the superior efficiency of compiled languages over interpreted ones, which have some overhead due to interpreting the code at runtime. Naz and Furia \cite{nanz2015comparative} evaluate Python and other 7 programming languages according to their performance, size of executable, conciseness, and failure proneness. The programming languages are evaluated using the Rosetta Code repository, which includes 745 programming problems. % Despite using the compiled version of Python (\ie CPython),
The results confirm the performance limitations of Python, which is instead praised for its conciseness.

The limited efficiency of Python is attributed to both its specification and implementation. The specification encompasses its syntax and semantics, while CPython, the reference implementation, is responsible for executing the code. Dynamic typing, a feature of the specification, increases accessibility but can also lead to slower runtime performance \cite{zhang2022quantifying}. A well-known bottleneck in CPython is the Global Interpreter Lock (GIL), which limits execution to a single thread within a process.

% Compilation as an approach
Compilation is frequently used by practitioners to combine the accessibility of Python with improved code efficiency. Just-In-Time (JIT) compilers convert Python code to machine code at runtime \cite{zhang2024python}. Numba \cite{lam2015numba} and PyPy \cite{PyPy2024} are two well-known Python JIT compilers. Ahead-Of-Time (AOT) compilation, instead, happens before running the code and usually generates an executable \cite{thom2018survey}. Nuitka \cite{Nuitka2024} is an AOT Python compiler that converts Python into optimized C or C++ code, compiles it into machine code, and generates an executable file. Some compilers optimize CPython directly, while others use a subset or a different implementation of Python. For example, PyPy is based on a subset of Python called Restricted Python (RPython). %Additionally, Python compilers can be domain-specific, such as Codon \cite{shajii2023codon}, an AOT compiler designed for scientific computing and numerical algorithms. 

% Novelty: lack of existing analysis on energy efficiency and more control over the characteristics of subjects 
% and the hardware platform
The \textbf{goal} of this work is to compare the efficiency benefits introduced by JIT and AOT Python compilers. We design an experiment that involves eight compilers: PyPy, Nuitka, Cython, Codon, MyPyC \cite{mypyc}, Numba, Pyston-lite \cite{pyston-lite}, and the experimental JIT compiler integrated in Python3.13 \cite{python313jit} against CPython. We test each compiler on seven functions extracted from the Computer Language Benchmarks Game (CLBG) \cite{CLBG}. The functions represent well-known problems in scientific computing (\eg fasta, mandelbrot). To our knowledge, the literature misses a comprehensive comparison of the performance and energy efficiency benefits of adopting Python compilers. In particular, we compare energy consumption, execution time, memory, and Last-Level Cache (LLC) miss rate. We study the compilers on two different testbeds, a server and an NUC. Existing experiments often overlook the characteristics of the benchmarks and of the platforms on which they operate. Code features such as the use of third-party libraries (\eg NumPy) and multithreading can improve the performance and energy efficiency of Python code \cite{reya2023greenpy}. Additionally, running a program on multiple cores and dynamically scaling frequency at runtime can significantly impact the results of our experiment. Van Kempen \etal \cite{van2024s} emphasize that language implementation and specification, the number of active cores and their frequency, and memory activity quantified as LCC cache misses should be considered in performance and energy studies. For this reason, we select a set of single-threaded functions from the same benchmark (\ie CLBG) that do not use third-party libraries. We execute our experiment on a single core, fixing CPU frequency to avoid any influence on the measurements. 

% Contributions
The main \textbf{contribution} of our study is an experiment that compares compiled Python code and its outcomes. We analyze the data collected during the experiment and provide insights into the execution time, energy and memory usage, LLC miss rate of each compiler. Our discussion highlights the compilers that offer the best optimization, comments on the effort required to use them, and provides future research paths. Additionally, this work includes a replication package  \cite{repl-package} that contains the scripts necessary to execute the experiment, repeat the data analysis, and access the raw data from our measurements. 

% Relevance
This study is essential for users, as it provides insights to help them choose a compiler based on performance, energy efficiency, and ease of use. Practitioners and researchers can use our findings as a foundation to enhance existing Python compilers.
\section{Related Work}
\label{s:related}
Python is popular among developers for its versatility and accessibility, but languages like C and Rust outperform it in performance and energy efficiency \cite{pereira2021ranking}. Research indicates that the inefficiency of Python stems from its specification and the CPython implementation. Literature shows various approaches to enhance the efficiency of Python \cite{zhang2022regcpython, melanccon2023executable, reya2023greenpy, pfeiffer2024energy}. %Python compilers can enhance performance and energy efficiency with minimal code changes. However, comparisons can be influenced by confounding factors, including code characteristics like high-performance library usage and testbed features such as the number of active cores and frequency.

%Simon Portegies Zwart \cite{portegies2020ecological} shares his concerns about the adoption of Python in astrophysics, a domain that involves problems requiring days for a solution and powerful computing resources. The author shows the poor performance of the Python implementation of the n-body problem compared to other programming languages, such as Java and C++. Furthermore, the study indicates that using the Numba compiler can greatly speed up the execution time, exceeding that of Java and Ada. Augier \etal \cite{augier2021reducing} builds upon the work of Zwart, differentiating between AOT and JIT compilers. The authors compare Pythran \cite{Guelton2015Pythran}, PyPy, and Numba on the execution of a single-threaded n-body problem running on a single core. The analysis suggests that compilation boosts execution time and energy efficiency and that Pythran results the most beneficial among the three.

Simon Portegies Zwart \cite{portegies2020ecological} raises concerns about Python performance in astrophysics, particularly for n-body problem simulations, highlighting inefficiency compared to Java and C++. He notes that using the Numba compiler can enhance execution speed, even surpassing that of Java and Ada. Building on the work of Zwart, Augier \etal \cite{augier2021reducing} compare the performance of AOT and JIT compilers, specifically Pythran \cite{Guelton2015Pythran}, PyPy, and Numba for a single-threaded n-body problem. Their analysis reveals that compilation improves both execution time and energy efficiency, with Pythran yielding the best results among the three. Zhang \etal \cite{zhang2024python} compare six Python JIT compilers: PyPy, GraalPy \cite{graalpy}, Pyjion \cite{pyjion}, Pyston, Jython \cite{juneau2010definitive}, and IronPython \cite{ironpython}, against a custom JIT compiler developed by the authors, called comPyler. They evaluate these compilers based on their execution time and memory usage using the pyperformance benchmark \cite{pyperformance}. PyPy and GraalPy provide the best speed-ups but have compatibility issues with CPython and can cause memory growth in long executions. Pyston offers the best compatibility with CPython and a modest speed improvement. Jython and IronPython generally perform slower than CPython, while the  performance of Pyjion is inconsistent and sometimes lags behind CPython, although it is fully compatible. Shajii \etal \cite{shajii2023codon} introduces Codon, a Python AOT compiler designed for resource-intensive tasks. The authors benchmark the performance of Codon against CPython, PyPy, and C++, demonstrating speed improvements of over 100 times in some cases, using implementations that don't rely on external libraries. Akeret \etal \cite{akeret2015hope} present HOPE, a Python JIT compiler for numerical astrophysical computations that matches C++ performance. Implemented on a subset of Python for numerical tasks, HOPE is compared against CPython, Numba, Cython, Nuitka, PyPy, and C++ using seven benchmarks. The results indicate that HOPE outperforms CPython by a factor of 2.4 to 119, achieving performance close to C++. While other compilers also improve upon CPython, none match the performance of HOPE and C++ in certain cases, highlighting compatibility issues with some compilers like Nuitka. Banijamali \cite{pouyeh2024impact} assesses the performance and energy efficiency of Codon compared to CPython and C++ (compiled with Clang) using 11 benchmarks from Codon, CLBG, and Programming Language and Compiler Benchmarks \cite{PLCB} across three input sizes (small, medium, big). Codon outperforms CPython in energy efficiency and speed in all benchmarks, though C++ often exceeds the performance of Codon. Additionally, Codon has longer compile times than C++.

% Summary of main points and novelty
Differently from \cite{zhang2024python, augier2021reducing, akeret2015hope, shajii2023codon}, we study energy efficiency and other than performance. This aspect is shared only by the study of Banijamali \cite{pouyeh2024impact}. Related work use code taken from pyperformance \cite{zhang2024python, shajii2023codon}, which can use third-party libraries that can influence code performance (\eg Numpy) or employ parallelism. We control the characteristics of the benchmarks and the testbed used in our experiment. Indeed, we use only code taken from the CLBG, which is single-threaded, and free from third-party libraries. In addition, we control testbed features, such as the number of active CPUs and CPU frequency that can influence software energy efficiency and performance.
\section{Study Design}
\label{s:design}
This study quantifies efficiency improvements in energy consumption and performance from Python code compilation. We conduct a controlled experiment comparing eight Python compilers to CPython. Our first research question focuses on energy usage:

\begin{questions}
    \item \textit{\RQone}
\end{questions}

Python code is often seen as energy inefficient, consuming more energy than other languages for similar tasks \cite{pereira2021ranking}. This study explores how compilation can enhance the energy efficiency of Python and identifies compilers that contribute to these improvements. We compile seven benchmarks using eight different Python compilers, execute them, and measure energy consumption.

There is often a negative correlation between software performance and energy efficiency \cite{weber2023twins}. This relationship varies based on software behavior and testbed settings \cite{castor2024estimating}. Our second research question focuses on how compilers optimize performance, measured by execution time, memory usage, and LLC miss percentage.

\begin{questions}
\item[RQ2.] \textit{\RQtwo}
\end{questions}

Different programs can exploit underlying resources in various ways. For example, software that requires significant processing typically uses more CPU than memory. In contrast, software that reads and writes files or allocates memory dynamically tends to put more stress on memory \cite{drepper2007every}. Compilers are well-known for optimizing how software uses computing resources. Common compiler optimizations include techniques applied to loops (\eg loop unrolling), constants (\eg constant folding), and frequently executed code in JIT compilers \cite{saeditowards}. We investigate to what extent Python compilers optimize execution time, memory usage, and the percentage of LLC misses. We quantify memory usage as the Resident Set Size (RSS), which corresponds to the physical memory used by a process, excluding swap memory \cite{gregg2019bpf}. Furthermore, we monitor LLC miss percentage, as suggested by Van Kempen \etal \cite{van2024s}. A high LLC miss percentage indicates that the application is unable to find the requested information in the cache, necessitating a fetch from the memory. This operation can result in significant latency penalties, which may slow down the application. We profile CPU usage but do not discuss it in detail, since we allocate all the workload to a single core, we expect the CPU to be fully committed during execution.

\subsection{Subjects Selection}
\begin{table}[h]
    \centering
    \scriptsize{
    \caption{Subjects Summary} %AOT: Ahead-of-Time, JIT: Just-in-Time}
    \label{tab:subjects}
    \begin{tabular}{@{}p{1cm}p{0.65cm}p{6cm}@{}}%{lll}
        \toprule
        \textit{Subject} & \textit{Type} & \textit{Description} \\
        \midrule
        Nuitka & AOT & Translates full Python code to optimized C++  \\
        Cython & AOT & Converts Python to C/C++ \\
        MyPyC  & AOT & Compiles Python to C extensions \\
        Codon  & AOT & Compiles a subset of Python to native machine code \\
        PyPy &  JIT &  Uses a JIT compiler based on a subset of Python (RPython)\\
        Numba & JIT &  Compiles Python into optimized machine code at runtime \\
        Pyston-lite & JIT & Adds lightweight JIT optimizations to the standard CPython interpreter \\
        Python 3.13 & JIT & Experimental JIT for CPython, introduced in Python 3.13  \\
        CPython & Interpreter & The reference Python implementation \\
        \bottomrule
    \end{tabular}
    }
\end{table}
The experiment involves eight Python compilers, each one with distinct characteristics. The subjects are chosen based on their popularity, proven efficiency benefits, active development, and support for Python 3, the last version of Python at the time of writing. For example, we excluded Jython as it supports only Python 2.7. Table \ref{tab:subjects} briefly describes the subjects of our study. 

We select four AOT compilers: Nuitka, Cython, Mypyc, and Codon. \textit{Nuitka} converts Python code into optimized C++ executables. Its GitHub repository has over 12400 stars and 650 forks on GitHub \cite{Nuitka2024}. Nuitka demonstrates a 3.7 times improvement over CPython in the Pystone benchmark\footnote{\url{https://nuitka.net/user-documentation/performance.html}} and consistently outperforms Python \cite{shajii2019seq}. \textit{Cython} is an AOT compiler that translates Python into C code and is highly popular, with over 40 million monthly downloads on PyPI\footnote{\url{https://pypistats.org/packages/cython}}. Behnel et al. \cite{behnel2010cython} report a 40 times speed-up in solving differential equations using Cython compared to Python. \textit{Mypyc} compiles Python code into C extensions. It is built on CPython and includes features like compiling Python classes and using unboxed representations for integers and booleans. It can improve performance by 1.5 to 5 times compared to standard Python\footnote{\url{https://mypyc.readthedocs.io/en/latest/introduction.html}}. \textit{Codon} translates a subset of Python to native machine code \cite{shajii2023codon}. Its repository on GitHub has more than 15000 stars. By using Python syntax and a limited set of semantics, it avoids features like dynamic typing and the GIL that can introduce overhead. % Codon differs from CPython and addresses its limitations, such as the GIL, which imposes single-threading, resulting in significant performance improvements.

We chose PyPy, Numba, Pyston-lite, and the experimental compiler of Python 3.13 as JIT compilers. \textit{PyPy} is a popular (more than 1000 stars on GitHub) alternative Python interpreter based on Restricted Python (RPython). PyPy makes Python code up to \textit{2.8} times faster than CPython \cite{PyPy2024}. It excels with code entirely written in Python. Therefore, code that does not rely on lower-level libraries. PyPy developers suggest that Python code may consume less memory than its CPython counterparts. 

\textit{Numba} compiles Python code into optimized machine code \cite{lam2015numba}, and it is designed for resource-demanding software, such as scientific software. The GitHub repository of Numba has more than 10000 stars. Developers must use decorators to specify which code to compile, with the compiled code cached for reuse. %Key features include vectorization and static typing, which are particularly effective with GPUs. 
Numba-compiled code can achieve speeds comparable to C or FORTRAN for numerical and array-oriented tasks. \textit{Pyston-lite} is built on top of CPython, and it is a lightweight version of the Pyston compiler \cite{pyston-lite}, which is no longer maintained. Low overhead, quickening, and aggressive attribute caching are among the main features of Pyston-lite, which is proven to boost the performance of Python 3.8 by 10\%. The developers of Pyston-lite state that it gets 100x more downloads per day compared to Pyston\footnote{\url{https://blog.pyston.org}}. In October 2024, Python 3.13 was released with an \textit{experimental JIT compiler} \cite{python313jit} that uses a "copy-and-patch" technique to match code patterns with pre-compiled machine code templates. We included this compiler in our study due to its significance to the Python community and its ongoing development.
%In October 2024, Python 3.13 was released along with an \textit{experimental JIT compiler} \cite{python313jit}. This compiler uses a technique called "copy-and-patch," which looks for patterns in the code to match against pre-compiled machine code templates to fill with code-specific information, such as memory addresses. We included this compiler in our study subjects due to its relevance to the Python community, as it is an initiative from the maintainers of the language and due to its actively development.
% Summary
%To summarize, our subjects include compilers with different characteristics, such as different Python implementations, target languages, and compilation algorithms. Although some compilers already exhibit some performance gains, we want to provide a common ground where compilers are evaluated based solely on their ability to optimize code efficiency without the influence of underlying platforms or code-level optimizations.

\subsection{Experimental Variables}
The \textit{benchmarks} represent the primary independent variable for this study. We select seven programming problems from the CLBG implemented in Python. We select code from the same codebase to reduce bias that may arise from the experience levels of developers. The CLBG code is intended to be written by developers with similar levels of experience. More experienced developers tend to write more efficient code compared to junior developers. Additionally, we choose single-threaded code that does not include any third-party libraries (\eg NumPy), which are known to enhance the performance of Python significantly.

The \textit{testbed}, defined by the operating system and underlying hardware, significantly impacts software efficiency. The operating systems can control the CPU frequency the software uses to save energy or boost performance \cite{ibrahim2016governing}. Linux has six governors that can be set to change the CPU frequency scaling policy \cite{ibrahim2016governing}. In addition, the OS uses different algorithms to supply tasks to available cores that can prioritize performance and energy savings. In this experiment, we execute Python code on a single core at a fixed frequency to avoid any influence of the testbed on the measurements. Thus, \textit{CPU frequency} and \textit{cores} can be considered as fixed factors of the study. We replicate the experiment on two different testbeds, an Intel NUC and a server, to see if the performance and energy consumption results are consistent. We consider the testbed to be a blocking factor in this study.

The dependent variable is energy consumption in KiloJoules (KJ) for RQ1. For RQ2, we measure the execution time in minutes (min), Memory usage in megabytes (MB), and LLC misses in percentage (\%).

\subsection{Experiment Design}
The experiment employs a \textit{full factorial design} where we execute each compiler with each benchmark and testbed. We combine 9 execution modes, namely 8 compilers plus the reference interpreter of Python (\ie CPython) with 7 benchmarks, obtaining 63 treatments. The experiment is repeated on each testbed separately, as we treat it as the blocking factor of our study. Each treatment is conducted 15 times on the NUC and 10 times on the server, leading to 945 runs on the NUC and 630 on the server. This results in a combined total of \textit{1575} runs. We randomized the runs on each testbed to prevent treatment characteristics from affecting our results.

\subsection{Data Analysis}
The data analysis is designed according to the guidelines of Wohlin \etal \cite{wohlinexperimentation}. We assess the normality of each group using the Shapiro-Wilk test and evaluate it graphically through Q-Q plots. If the groups are normally distributed, we apply the ANOVA test. For non-normally distributed groups, we use the Kruskal-Wallis test. This approach helps us determine whether adopting different compilers affects energy consumption and performance. We anticipate a difference between the compiled and interpreted versions based on previous research. However, the magnitude of these differences is still unclear, as prior studies may have been affected by variables that we are controlling in this investigation, including CPU frequency, the number of processors, and developer experience. We use the effect size to check whether the difference is significant. We use Cohen's d test for normally distributed data and Cliff's Delta test for non-normal data.
\section{Experiment Execution}
\begin{figure}
    \centering
    \includegraphics[width=\columnwidth]{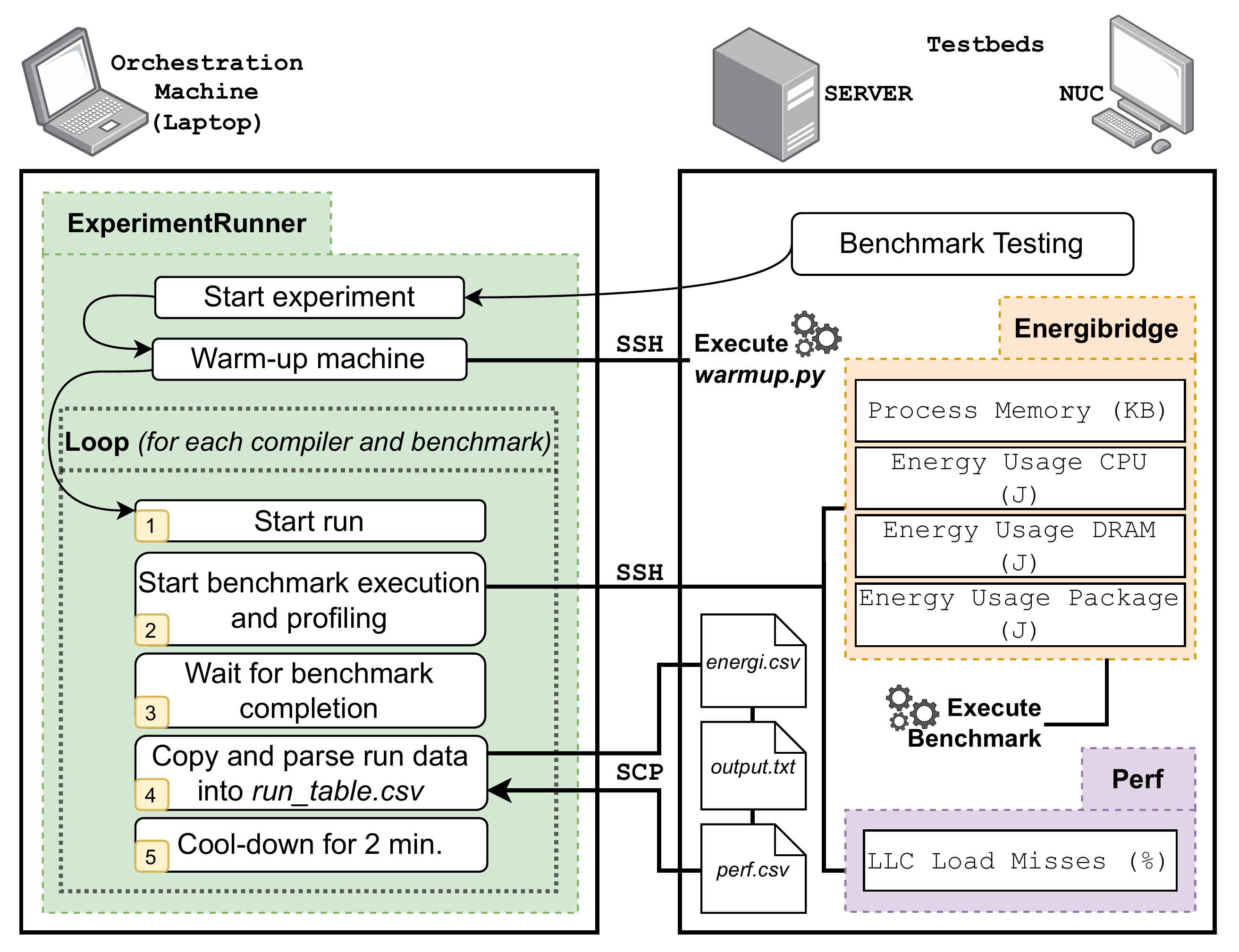}
    \caption{Experiment Execution}
    \label{fig:experiment_execution}
    \vspace{-0.6cm}
\end{figure}

The section outlines the execution of the experiment, including the experimental setup and measurement tools used. The experiment lasted 304 hours: 136 hours on the NUC and 168 hours on the server. Figure \ref{fig:experiment_execution} shows the all the steps involved in the experiment execution and the experimental setting. 

\subsection{Preparation}
Before the experiment was executed, we installed the compilers on both NUC and the server. We use Codon \texttt{0.17}, Numba \texttt{0.60}, Cython \texttt{3.0.11}, Nuitka \texttt{2.5.9}, Mypyc \texttt{1.14.1}, Pyston-lite \texttt{2.3.5}, PyPy \texttt{7.3.15}. All the compilers are executed using the same version of Python, namely Python \texttt{3.10}, except for the experimental JIT compiler, which uses Python \texttt{3.13}. The selected code is single-threaded and free of any third-party library code. We chose code from the CLBG that indicated at least one minute of execution time to collect enough data for our study. The code is executed using the input included on the CLBG page of each implementation and refined, in some cases, to be successfully compiled. In particular, we needed to ensure compatibility with Codon and Numba, primarily due to differences in how these compilers handle Python objects and types and printing. Numba and Codon code required class types to be specified. In addition, Numba code needed the \texttt{@njit} decorator to each function to enforce function compilation. We minimized code changes, ensuring the control and data flow remained intact. We created a set of tests for each benchmark and compiler and compared their output with the output of the interpreted source code to ensure their correctness. The programs for AOT compilers are built using a series of build scripts. The code, test cases, and build scripts can be found in the replication package of the study \cite{repl-package}.

We control some features of the testbed. We disable Intel Hyper-Threading \cite{tau2002empirical}, which allows the execution of multiple threads on a core, from the BIOS on both machines. We fix the frequency of each CPU through Linux governors. The latter regulates CPU frequency scaling and can be set with a parameter of the kernel. We fix the frequency on the server to 1.6 GHz. We use the powersave governor on the NUC, which kept its frequency to its minimum (\ie 2.1Ghz). We deactivate any active background processes, which means the termination or temporary suspension of non-essential processes (\eg Docker).

\subsection{Experimental Setting}
As depicted in Figure \ref{fig:experiment_execution}, our experimental setting involves two machines: an orchestration machine and two testbeds, namely a server and an NUC. The orchestration machine arranges all the steps of experiment execution, such as starting the measurements tool, the benchmarks, and collecting measurements. We use a laptop to orchestrate the experiment, and it has an Intel Core i7-9750H with six physical cores, 16 GB of RAM, and 512 GB SSD running Ubuntu 20.04. The testbeds execute the benchmarks for each subject and run the measurement tools in the background. We repeat the same experiment on each testbed. The server has an Intel Xeon E3-1231 CPU with four physical cores, 32 GB of RAM, and a 1 TB hard drive. The NUC has an Intel Core i7-1260P with 12 physical cores, 32 GB of RAM, and a 512 GB hard drive. Both testbeds run Ubuntu 24.0 as the operating system.

We use EnergiBridge, the \texttt{time} Python module\footnote{\url{https://docs.python.org/3/library/time.html}}, and \texttt{perf} \cite{perf} as our measurement tools. EnergiBridge \cite{energibridge} is a cross-platform measurement utility that supports Linux, Windows, and MacOS, along with Intel, AMD, and Apple ARM CPU architectures. We utilize it to track energy and memory usage at fixed intervals of 200 milliseconds. EnergiBridge uses the RAPL interface provided with Intel CPUs to profile energy usage. RAPL provides the energy usage of the DRAM, cores, and uncore components (\eg LLC and memory controller). It is important to note that Intel CPUs from the 11th generation onward have removed DRAM energy readings from the RAPL domain on non-server-grade processors. Therefore, we could not obtain the energy usage of DRAM in our experiment on the NUC. We use \texttt{time} to monitor the execution time of benchmark executions. We call \texttt{time} before and after benchmark execution in the orchestration scripts. We calculate the execution time as $end\_time - start\_time$. We employ \texttt{perf} to measure the LLC miss percentage. The \texttt{perf} tool provides insights into system performance by collecting real-time kernel events and its specific to Linux.

\subsection{Execution}
Figure \ref{fig:experiment_execution} outlines the experiment execution procedure. All steps are coordinated using the Experiment Runner \cite{experiment-runner}, a framework designed to automate the various phases of the experiment on an orchestration machine, typically a laptop. This framework allows users to define the operations to be performed during each phase of the experiment and to establish their sequence. Communication between the laptop and the testbeds occurs through an SSH connection, which is implemented using the Paramiko Python plugin \cite{paramiko}. We utilize Paramiko to establish the SSH connection, automate command execution on the remote machines, and collect data from the measurement tools. All commands executed on the testbeds, such as warm-up, benchmark execution, and measurement tools, are run on the first core of the processor. We ensure these commands are tied to the first core by using the command \texttt{taskset -c 0}.
%
% Before the execution of the experiment, we warm up each testbed for 2 minutes using a Python implementation of the Fibonacci sequence to bring the machine to a stable temperature and avoid the influence of a cold start. Afterward, we create all the commands and directories to start and store the measurements for a specific run. We start EnergiBridge to profile energy and memory usage and the time tool for execution time. EnergiBridge is started asynchronously, and once it starts, it outputs the Process ID (PID) of the process corresponding to the execution of a benchmark. The PID is given to the perf tool that tracks the LLC miss rate. The orchestration machine waits until the execution of the process is over. The generated output files from the measurement tools and the script output are then copied from the testbeds machines to the orchestration machine. The \texttt{output.txt} file is then verified to ensure the validity of the run, and then the \texttt{perf.csv} and \texttt{energibridge.csv} files are processed to store the measurements in the \texttt{run\_table.csv} file. Lastly, the experiment is paused for 2 minutes for the experimental machines to cool down.
We adhere to the guidelines typically used for experiments on software energy consumption \cite{cruz2021green, malavolta2024ten}. Before executing the experiments, we warm up each testbed for 2 minutes by running a CPU-intensive task (such as calculating the Fibonacci sequence) to stabilize the temperature, as energy consumption is significantly influenced by the hardware temperature \cite{cruz2021green}. Following this, we create directories to organize and store the outputs from the executions and their measurements. Each run is tracked using the Process ID (PID) generated at the start of the execution. This PID is then passed to the measurement tools, which operate in the background, allowing for the profiling of the specific execution. The output of each execution is saved in a file named `output.txt`, and the corresponding measurements are collected by the orchestration machine.  At the end of each execution, we pause the experiment for an additional 2 minutes to allow the heat accumulated during the run to dissipate.

% verify the validity of the \texttt{output.txt} file, and process the \texttt{perf.csv} and \texttt{energibridge.csv} files into \texttt{run\_table.csv}. Finally, we paused the experiment for 2 minutes to allow the machines to cool down.
\section{Results}
\label{s:results}
This section presents the results of our experiment. Table \ref{tab:platform_stats} presents the descriptive statistics of the data collected on the server, summarizing the statistics obtained by aggregating the measurements for each benchmark.

% MAIN TABLE
\begin{table*}
\centering
\caption{Descriptive Statistics from Server data. The highlighted number shows the minimum average value.}
\label{tab:platform_stats}
\scriptsize{
\begin{tabular}{lccccc|ccccc|ccccc|ccccc}
\toprule
\textit{Subject} & \multicolumn{5}{c}{\textit{Energy Consumption (KJ)}} & \multicolumn{5}{c}{\textit{Execution Time (min)}} & \multicolumn{5}{c}{\textit{Memory Usage (MB)}} & \multicolumn{5}{c}{\textit{LLC Load Misses (\%)}} \\
 & mean & std & min & 50\% & max & mean & std & min & 50\% & max & mean & std & min & 50\% & max & mean & std & min & 50\% & max \\
\midrule
CPython & 16.41 & 12.97 & 5.29 & 11.15 & 46.06 & 24.98 & 19.95 & 8.16 & 17.27 & 71.53 & 3.56 & 2.85 & 0.00 & 3.90 & 7.21 & 21.73 & 24.49 & 0.61 & 12.91 & 74.48 \\
PyPy & 1.54 & 0.89 & 0.36 & 1.60 & 3.04 & 2.36 & 1.39 & 0.58 & 2.25 & 4.77 & 5.18 & 6.97 & 0.00 & 1.01 & 27.27 & \cellcolor{customblue!30} 13.25 & 14.14 & 0.15 & 5.53 & 36.93 \\
Numba & 1.33 & 0.76 & 0.23 & 1.08 & 2.69 & 2.15 & 1.28 & 0.38 & 1.74 & 4.52 & 3.29 & 2.72 & 0.00 & 1.56 & 7.23 & 18.22 & 14.53 & 3.52 & 15.30 & 53.30 \\
Pyston-lite & 16.87 & 13.45 & 5.24 & 11.12 & 46.38 & 25.85 & 20.83 & 8.15 & 17.19 & 71.62 & 2.78 & 2.69 & 0.00 & 1.01 & 7.12 & 20.68 & 24.50 & 0.41 & 10.54 & 74.12 \\
Python 3.13 JIT & 12.57 & 7.43 & 5.79 & 10.29 & 27.39 & 19.41 & 11.62 & 8.93 & 15.66 & 42.41 & 3.88 & 3.11 & 0.00 & 2.34 & 8.80 & 18.13 & 24.82 & 0.33 & 5.68 & 73.87 \\
Nuitka & 14.32 & 9.75 & 4.55 & 13.69 & 33.89 & 21.88 & 14.95 & 7.05 & 21.87 & 52.56 & \cellcolor{customblue!30} 1.75 & 0.98 & 0.00 & 1.71 & 4.27 & 58.73 & 12.07 & 22.77 & 63.80 & 74.22 \\
Cython & 14.93 & 9.57 & 4.75 & 15.50 & 33.59 & 22.80 & 14.71 & 7.40 & 24.46 & 51.95 & 2.84 & 2.55 & 0.00 & 1.29 & 7.54 & 20.38 & 24.47 & 0.36 & 9.35 & 74.46 \\
Codon & \cellcolor{customblue!30} 0.64 & 0.49 & 0.10 & 0.41 & 1.47 & \cellcolor{customblue!30} 1.04 & 0.81 & 0.19 & 0.58 & 2.40 & 2.35 & 2.12 & 0.00 & 1.01 & 7.58 & 20.89 & 12.38 & 3.06 & 18.56 & 44.57 \\
Mypyc & 15.09 & 12.93 & 5.69 & 8.57 & 53.53 & 22.73 & 19.82 & 8.86 & 13.24 & 85.00 & 3.15 & 2.75 & 0.00 & 1.05 & 7.22 & 20.06 & 22.01 & 0.37 & 13.00 & 66.84 \\
\bottomrule

\end{tabular}
}
\end{table*}

% MAIN FIGURE
\begin{figure*}
    \centering
    \begin{subfigure}{\textwidth}
        \centering
        \includegraphics[width=\textwidth]{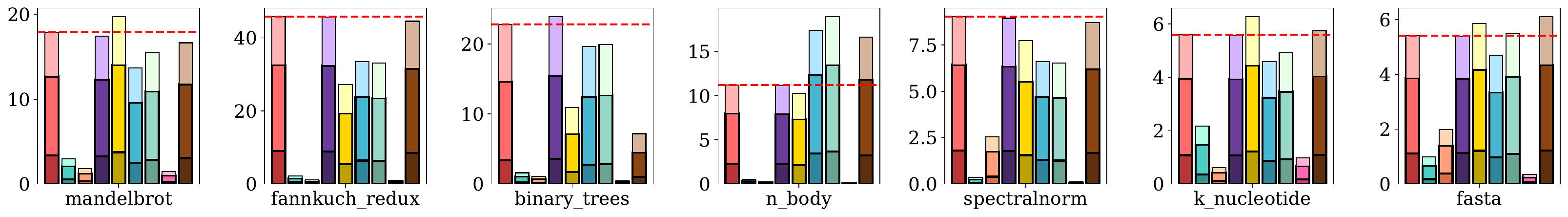}
        \caption{Average Energy Consumption (KJ) spent on the Server for each benchmark. The bars are colored using a gradient scale representing the energy consumed by the core, uncore components, and DRAM. The gradient ranges from light (DRAM) through intermediate (uncore components) to dark (core).}
        \label{fig:gl2_energy}
    \end{subfigure}
    \vspace{0.2cm}
    \begin{subfigure}{\textwidth}
        \centering
        \includegraphics[width=\textwidth]{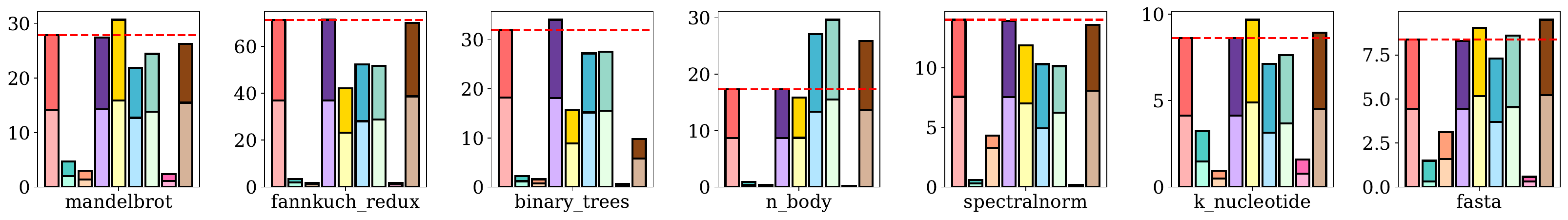}
        \caption{Average Execution Time (in minutes) for each benchmark. The bars feature two different shades: the lighter color represents the execution time on the NUC, while the darker color indicates the execution time on the server.}
        \label{fig:time_bars}
    \end{subfigure}
    \vspace{0.2cm}
    \begin{subfigure}{\textwidth}
        \centering
        \includegraphics[width=\textwidth]{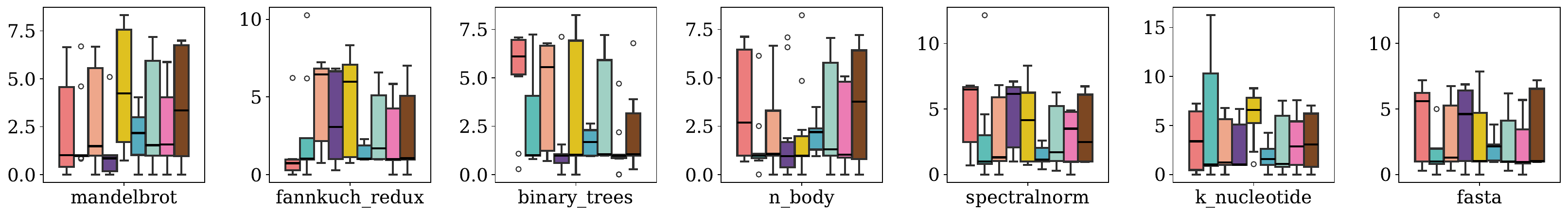}
        \caption{Memory Usage (in megabytes) for each benchmark executed on the server.}
        \label{fig:mem_box}
    \end{subfigure}
    \caption{
        Average energy usage, execution time, and memory usage for each benchmark by compiler. The dashed red line represents the threshold of the CPython implementation for a given benchmark.\\
        Legend:
        %     ['cpython', 'pypy', 'numba', 'pyston-lite', 'py3.13-jit', 'nuitka', 'cython', 'codon', 'mypyc']
        \protect\tikz\protect\draw[fill={rgb,255:red,255;green,107;blue,107}] (0,0) rectangle (0.5em,0.5em); CPython,
        \protect\tikz\protect\draw[fill={rgb,255:red,78;green,205;blue,196}] (0,0) rectangle (0.5em,0.5em); PyPy,
        \protect\tikz\protect\draw[fill={rgb,255:red,255;green,160;blue,122}] (0,0) rectangle (0.5em,0.5em); Numba,
        \protect\tikz\protect\draw[fill={rgb,255:red,106;green,61;blue,154}] (0,0) rectangle (0.5em,0.5em); Pyston-lite,
        \protect\tikz\protect\draw[fill={rgb,255:red,255;green,215;blue,0}] (0,0) rectangle (0.5em,0.5em); Python 3.13 JIT,
        \protect\tikz\protect\draw[fill={rgb,255:red,69;green,183;blue,209}] (0,0) rectangle (0.5em,0.5em); Nuitka,
        \protect\tikz\protect\draw[fill={rgb,255:red,152;green,216;blue,200}] (0,0) rectangle (0.5em,0.5em); Cython,
        \protect\tikz\protect\draw[fill={rgb,255:red,255;green,105;blue,180}] (0,0) rectangle (0.5em,0.5em); Codon,
        \protect\tikz\protect\draw[fill={rgb,255:red,139;green,69;blue,19}] (0,0) rectangle (0.5em,0.5em); MyPyc
    }
    \label{fig:combined}
\end{figure*}

\subsection{RQ1: \RQone}
\label{ss:rq1}
% Descriptive statistics
Table \ref{tab:platform_stats} shows the average energy consumption in KiloJoules (KJ) for each subject running on the server. The mean is obtained by aggregating the data collected for each benchmark. We notice that CPython and Pyston-lite present the highest energy consumption value, suggesting that compilation can boost the energy efficiency of interpreted Python code. It is worth noting that the average energy consumption drops when executing the code using PyPy, Numba, and Codon. Codon shows the smallest average energy consumption value. However, energy optimization is inconsistent across benchmarks and seems highly influenced by the characteristics of the code. Figure \ref{fig:gl2_energy} describes the average energy consumption value across benchmarks and compilers. We can see that, in some cases, compilation can increase the energy consumption of Python code. For example, this case happens when compiling n\_body with Nuitka, Cython, and Mypyc. We formally check whether there is a difference between data collected from varying subjects using the Kruskal-Wallis test. We aggregate the data collected for each benchmark to perform the test. The Shapiro-Wilk test results are way below our significance level of 0.05 due to the significantly different scale of the energy values retrieved for each benchmark. This observation is confirmed by Figure \ref{fig:gl2_energy}. The Kruskal-Wallis test shows that there is a significant difference between the groups. We use Cliff's Delta to compare aggregated CPython data against the compiled code data. The effect size is large (1.0) for PyPy, Numba, and Codon, while in the other cases is negligible.

The NUC used in this study does not provide the DRAM measurements as the processor of the NUC lacks support for the RAPL DRAM domain. Therefore, we analyzed the energy consumed by the CPU, namely the core and the other components of the CPU package. The experiment performed on the NUC presents the same results as the one done on the server. PyPy, Numba, and Codon present the smallest average energy consumption, while CPython and Pyston-lite have the highest. Additionally, the energy varies across benchmarks, showing that even on the NUC, the compilers increase the energy consumption in the same cases experienced on the server. The Cliff's Delta confirms the results obtained for the server. Thus, there is a large effect size for PyPy, Numba, and Codon and a negligible effect for the rest of the compilers. The replication package provides the complete data analysis for both NUC and server experiments \cite{repl-package}.

\subsection{RQ2: \RQtwo}
\label{ss:rq2}

\subsubsection{Execution Time}
The results obtained for the execution time reflect those of RQ1. Table \ref{tab:platform_stats} reports the execution time (in minutes) for each subject on the server. We notice that, as for energy consumption, PyPy, Numba, and Codon stand out for their short average execution time. There is a difference of 23.94 minutes between CPython and Codon. The execution time difference between CPython and the compilers changes across benchmarks but is consistent in both the NUC and server. Figure \ref{fig:time_bars} compares the average execution time obtained using CPython with the compilers for each benchmark. The compilers can speed up most of the benchmarks with some exceptions, such as n\_body compiled with Nuitka, Cython, and MyPyc.  Figure \ref{fig:time_bars} confirms the significant improvement obtained with PyPy, Numba, and Codon.

% aligment nuc, gl2, and energy
Our results indicate that the energy consumption and execution time data obtained from the NUC and the server are consistent. Figure \ref{fig:time_bars} displays the execution time for each benchmark and subject derived from the experiments conducted on the NUC. When we compare Figure \ref{fig:time_bars} and Figure \ref{fig:gl2_energy}, we observe that both figures exhibit a similar pattern of improvement or detriment for each subject and benchmark. However, the average execution time values differ, which is expected given that we performed the experiment on two different machines. The same reasoning applies when comparing the execution time and energy consumption data. This observation suggests that there may be a linear relationship between energy consumption and execution time in our study.

% formal check
The Shapiro-Wilk test conducted on the data aggregated for each subject yields p-values significantly below our significance level of 0.05. We then use the Kruskal-Wallis test to determine if there are differences in execution time among the groups. This test also returns a p-value less than 0.05, indicating the presence of a difference. When using Cliff's Delta to compare CPython with each compiler group, we find a large effect size (1.0) for PyPy, Numba, and Codon, while the remaining compilers show a small and negligible effect size.

\subsubsection{Memory Usage}

Table \ref{tab:platform_stats} presents the average memory usage in megabytes for each compiler on the server. PyPy exhibits the highest average memory usage. Nuitka uses the least average memory. We notice that the memory usage presents high variability if we inspect the data for each benchmark. Figure \ref{fig:mem_box} includes a boxplot for each benchmark, where each box represents the data collected for a specific compiler on the server. We see that the boxes are overlapping in every subplot. Therefore, we are not able to identify any pattern in the data. We cannot confidently point to a compiler that improves the memory usage of CPython. The measurements taken on the NUC show a similar trend. The data presenting the least variability comes from the executions done with Nuitka, as also shown by the data sorted per subject in Table \ref{tab:platform_stats}. Our memory measurement ranges from a maximum of a few kilobytes to over 5 megabytes, reaching a peak of 11 megabytes during the execution of k\_nucleotide with PyPy on the server. 

The data collected for each subject does not follow a normal distribution, as indicated by the Shapiro-Wilk test. The Kruskal-Wallis test suggests that there are differences among the groups. These differences are reflected in Cliff's Delta effect size, which shows a small positive impact for both Nuitka and Codon, while the effect size for the other subjects is negligible. For the NUC, the test reveals a small positive effect size for Nuitka, Codon, and PyPy, whereas the other compilers show a negligible effect size.

% All the compilers demonstrate a significant reduction in memory usage. Among them, Codon, MyPyc, and Cython report the lowest memory usage. Furthermore, memory consumption varies significantly across different testbeds. Figure \ref{fig:memory_bars} compares the average memory usage percentage across benchmarks, highlighting the data collected on the NUC with a brighter color and the server data with a darker shade. All the compilers demonstrate a reduction in memory usage, although the extent of this reduction varies depending on the benchmark used. However, this is not the case for Python-lite and Python 3.13 JIT. Compiling on the NUC does not significantly affect the average memory usage of Python code. Only binary\_trees and k\_nucleotide show some variability on the NUC. For binary\_trees, PyPy, Python 3.13-JIT, and Codon introduce a reduction of memory usage.

%The Kruskal-Wallis test executed using the data collected on the NUC suggests a difference between the groups. The Cliff's Delta, instead, confirms what is shown in Figure \ref{fig:memory_bars}. The test returns a small negligible effect size for all the compilers except for Numba, which has a medium effect size of -0.39. This means that Numba increases the average memory usage. When observing Figure \ref{fig:memory_bars}, we see that this result may be influenced by the execution of binary\_trees and k\_nucleotide as they are the only cases in which memory usage varies significantly on the NUC.

\subsubsection{Cache usage}
We analyze the percentage of LLC misses to determine if it affects the execution time of Python code. When an LLC miss occurs, the software must retrieve data from DRAM, which can lead to slower execution. Table \ref{tab:platform_stats} displays the LLC load miss rates observed on the server. PyPy demonstrates the lowest average LLC miss rate at 13.25\%, followed by the Py3.13 JIT compiler at 18.13\% and Numba at 18.22\%. In contrast, CPython and the other compilers have an average LLC miss percentage exceeding 20\%. Notably, Nuitka has the highest LLC miss rate at 58.73\%.
% The coefficient of variation is high (above 1) for all subjects except for Codon, Numba, and Nuitka, indicating significant variability in the data.This suggests that a lower LLC miss rate may not be a reliable indicator of improvement. 
The LLC miss rates vary significantly across different benchmarks and are not consistently lower than the miss rate of CPython. However, Nuitka consistently shows a value above 50\% in each case. There are some cases where the miss rate of CPython is lower than other compilers, such as spectralnorm and fasta. The Kruskal-Wallis test indicates a significant difference between the groups on the server. The Cliff's Delta confirms the observations made using descriptive statistics. On the server, Nuitka has a significantly larger negative effect size (-0.70) than CPython. The other compilers result in a small and negligible effect size. Looking at the data obtained on the server, in our setting, none of the compilers could decrease the cache miss rate sensitively.

% NUC
This result is confirmed by the NUC data. The experiment conducted on the NUC confirms a consistently high miss rate for Nuitka, which stands at 68.92\%. Following Nuitka, Numba exhibits a miss rate of 24.17\%. In contrast, CPython and the other compilers show lower miss rates, ranging from 12.26\% to 13.47\%. Notably, the experimental Python 3.13 JIT compiler has the smallest miss rate among them. Therefore, the percentage of LLC misses varies significantly across the two testbeds. The Kruskal-Wallis test suggests differences between the groups. The Cliff's Delta executed with the NUC data shows a large negative effect size for Numba, Nuitka, PyPy, and Codon. For the remaining compilers, the difference is negligible.
\section{Discussion}
\label{s:discussion}
This section describes the results of our experiment for each dependent variable and provides insights for the practitioners and future research directions.
\begin{figure}[h]
    \centering
    \begin{subfigure}{\columnwidth}
        \centering
        \includegraphics[width=\columnwidth]{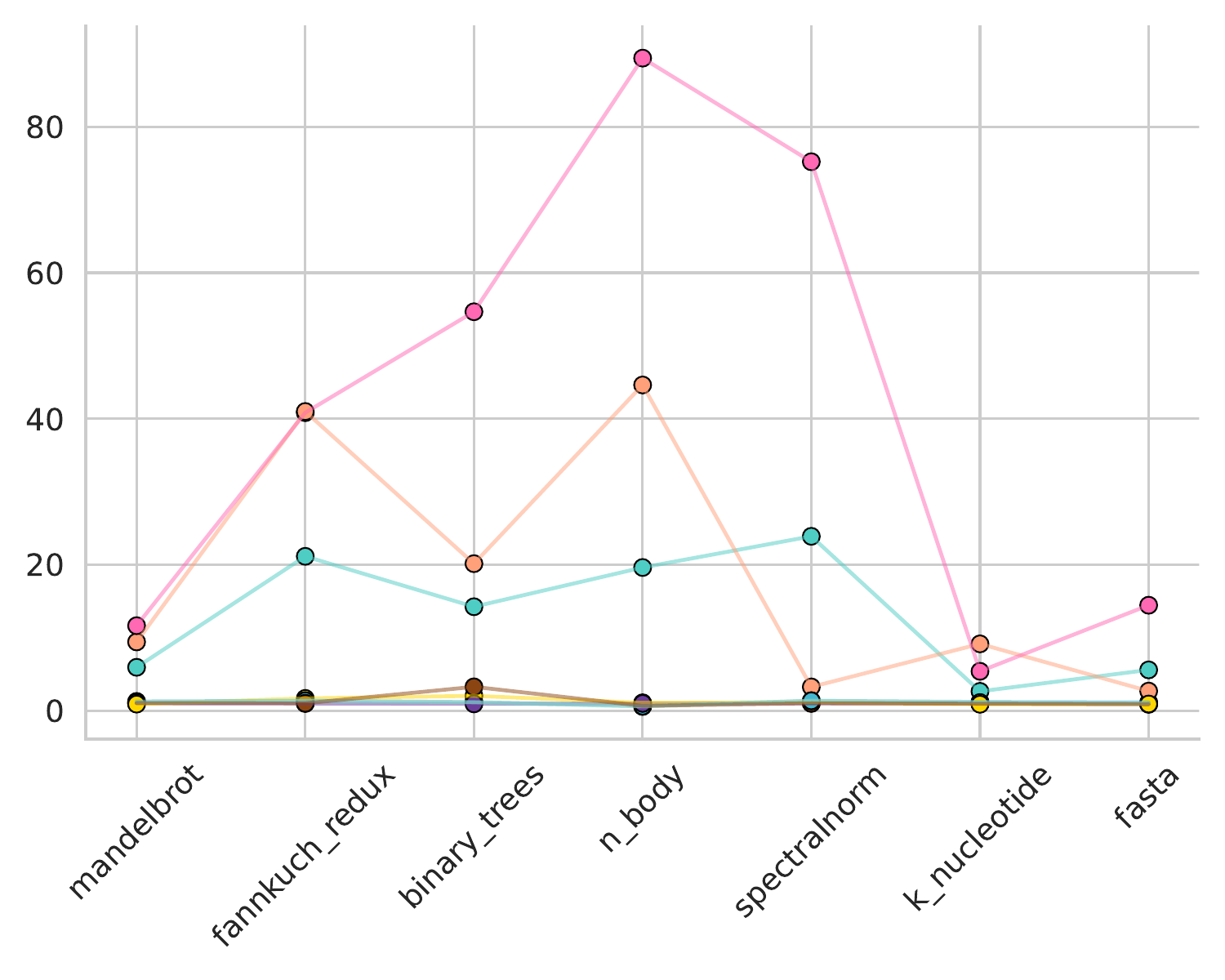}
        \caption{Speedup.}
        \label{fig:speedup}
    \end{subfigure}
    \begin{subfigure}{\columnwidth}
        \centering
        \includegraphics[width=\columnwidth]{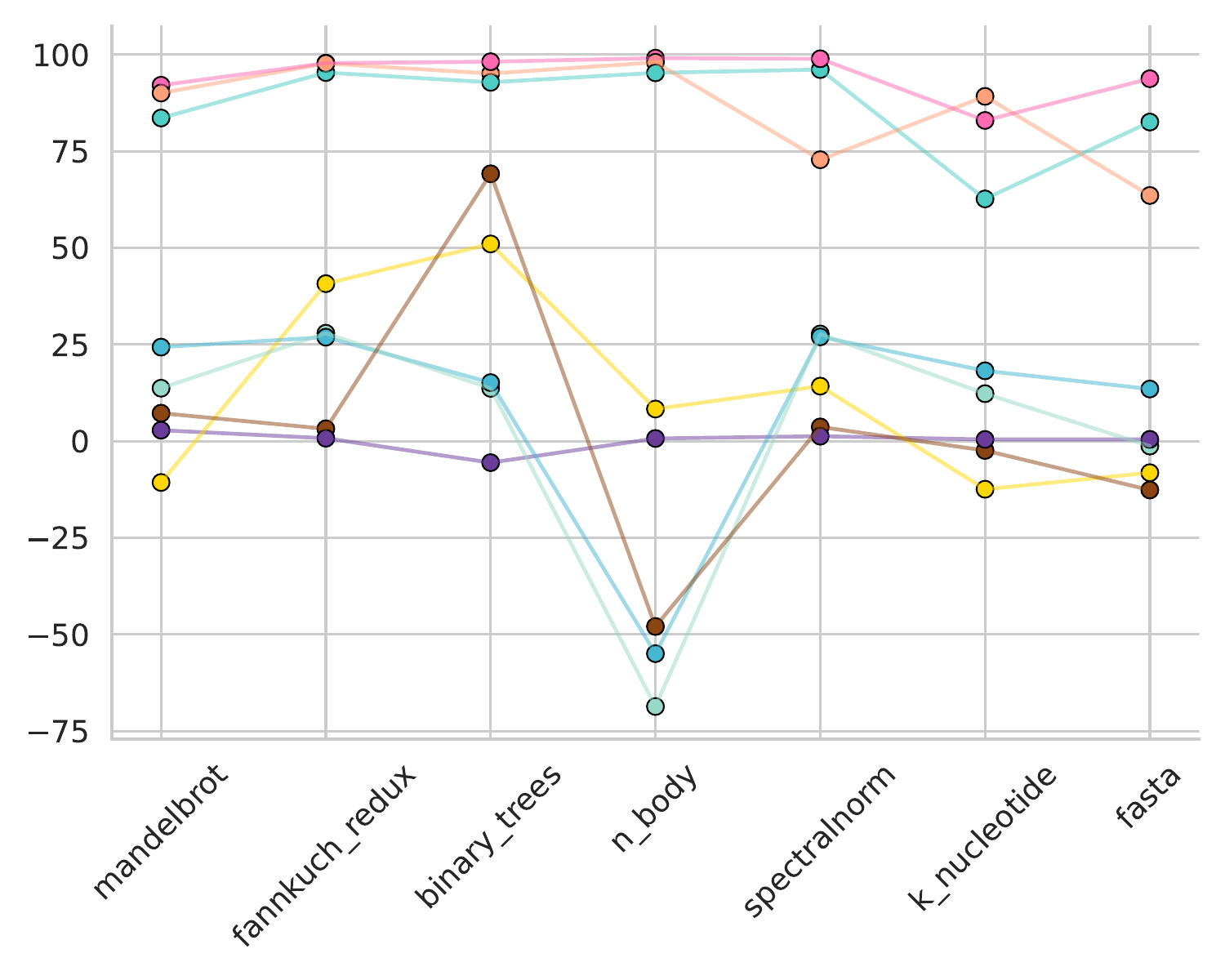}
        \caption{Energy Improvement (\%).}
        \label{fig:energy_improvement}
    \end{subfigure}
    %\begin{subfigure}{\columnwidth}
    %    \centering
    %    \includegraphics[width=\columnwidth]{figs/GL2_PROCESS_MEMORY_improvement.pdf}
    %    \caption{Memory usage improvement (\%).}
    %    \label{fig:memory_improvement}
    %\end{subfigure}
    \caption{Speedup and Energy improvement on the server across benchmarks compared to CPython.
        Legend: 
        %     ['cpython', 'pypy', 'numba', 'pyston-lite', 'py3.13-jit', 'nuitka', 'cython', 'codon', 'mypyc']
        %\protect\tikz\protect\draw[fill={rgb,255:red,255;green,107;blue,107}] (0,0) rectangle (0.5em,0.5em); CPython,
        \protect\tikz\protect\draw[fill={rgb,255:red,78;green,205;blue,196}] (0,0) rectangle (0.5em,0.5em); PyPy,
        \protect\tikz\protect\draw[fill={rgb,255:red,255;green,160;blue,122}] (0,0) rectangle (0.5em,0.5em); Numba,
        \protect\tikz\protect\draw[fill={rgb,255:red,106;green,61;blue,154}] (0,0) rectangle (0.5em,0.5em); Pyston-lite,
        \protect\tikz\protect\draw[fill={rgb,255:red,255;green,215;blue,0}] (0,0) rectangle (0.5em,0.5em); Python 3.13 JIT,
        \protect\tikz\protect\draw[fill={rgb,255:red,69;green,183;blue,209}] (0,0) rectangle (0.5em,0.5em); Nuitka,
        \protect\tikz\protect\draw[fill={rgb,255:red,152;green,216;blue,200}] (0,0) rectangle (0.5em,0.5em); Cython,
       \protect\tikz\protect\draw[fill={rgb,255:red,255;green,105;blue,180}] (0,0) rectangle (0.5em,0.5em); Codon,
       \protect\tikz\protect\draw[fill={rgb,255:red,139;green,69;blue,19}] (0,0) rectangle (0.5em,0.5em); MyPyc   
    }
    \label{fig:improvement}
    \vspace{-0.5cm}
\end{figure}

\subsection{Considerations on Energy Consumption}
The results for the RQ1 show that compilation can significantly improve the energy efficiency of Python code. In particular, PyPy, Numba, and Codon present significant improvements, while for other compilers the impact is negligible. The benefits of compilation vary according to the code at hand, where in some cases compilation is detrimental for energy efficiency.  Figure \ref{fig:energy_improvement} shows the percentage improvement in energy efficiency across benchmarks for each compiler with respect to CPython. We calculate the values in Figure \ref{fig:energy_improvement} using the data collected on the server. We notice that the improvement of PyPy, Numba, and Codon is consistent across benchmarks, where the smallest improvement is 62.64\% when k\_nucleotide is executed with PyPy. In contrast, the highest improvement is provided by n\_body compiled with Codon (\ie 99.06\%). Overall, PyPy, Numba, and Codon provide 86.89\%, 86.61\%, and 94.66\% energy improvement.

% R2: code characteristics matter
Despite our results presenting a negligible effect size for the remaining compilers, we can see that all compilers can optimize most benchmarks. The compilation results are highly disadvantageous for n\_body, when compiled with Mypyc, Cython, and Nuitka. This case affected the magnitude of the effect size analysis using aggregated benchmark data in Section \ref{ss:rq1}. Mypyc, Cython, and Nuitka increase energy consumption of 47.96\%, -68.65\%, -54.98\%, respectively, for n\_body. These compilers are all AOT compilers that convert Python to C/C++. We hypothesize that the source of inefficiency is the significant amount of mathematical operations performed by n\_body on Python lists. Indeed, Mypyc, Cython, and Nuitka may need to adapt Python features like type inference, list operations, and type to C and C++, which do not support these features.

% R3: cache role in software energy efficiency
On the server, we collect all the RAPL domains supported by the server, namely the package, the core, and the DRAM. The package domains contain the core and uncore components, such as the LLC and the memory controller. Figure \ref{fig:gl2_energy} presents the above-mentioned domains stacked in a bar chart for each benchmark and compiler. The bright gradient of the bar color represents the average energy consumed by the DRAM, while the intermediate and the dark colors show the uncore components and core consumption. The uncore components consume the most energy across benchmarks and compilers, followed by the core and the DRAM. We think that the high energy consumption can be attributed to LCC usage.  We suggest further investigation into the impact of LLC on software energy consumption.

\begin{tcolorbox}%[
%  colback=customblue!10,
%  colframe=customblue!80!black,
%  coltitle=white,
%  colbacktitle=customblue!80!black,
%  fonttitle=\bfseries,
%  left=1mm,
%  right=1mm,
%  top=1mm,
%  bottom=1mm,
%  title=Summary
%]
\textbf{\textit{Summary}} - Compilation greatly enhances the energy efficiency of interpreted Python code, with varying impacts based on code characteristics. Codon leads with an average energy improvement of 94.66\%, followed by PyPy (86.89\%) and Numba (86.61.\%) on the server. Most energy consumption comes from uncore components like the Last-Level Cache, memory controllers, and interconnect.
\end{tcolorbox}

\subsection{Considerations on Execution Time}
% results
Execution time can be significantly reduced by compiling Python code. Compilers like Codon, PyPy, and Numba provide consistent and substantial speed improvements across various benchmarks, as illustrated in Figure \ref{fig:time_bars}. If we calculate the percentage improvement, we get that Codon offers an impressive improvement of 94.18\%, followed by PyPy at 86.67\% and Numba at 85.86\%. The impact of the other compilers results is small or negligible. This observation is more evident if we consider the speedup calculated as $\bar x_{cpython}/\bar x_{compiler}$ where $\bar x_{i}$ is the average execution time obtained executing Python code using $i$, which can be CPython or a compiler in our case. Figure \ref{fig:speedup} shows the speedup across benchmarks executed on the server. The figure supports our findings, demonstrating significant and consistent speed improvements with Codon, Numba, and PyPy. The n\_body benchmark runs approximately 89 times faster when executed with Codon than CPython. Additionally, both Numba and PyPy show impressive speed boosts, with Numba achieving a speedup of 44 times on the n\_body benchmark and PyPy making spectralnorm 23 times faster.  

% slowdowns
The compilers tested in this study offer overall improvements, even if they sometimes introduce slowdowns. For example, the execution time of fasta and k\_nucleotide is longer when using the experimental Python 3.13 JIT compiler. This seems to be due to the compilation method and code characteristics, which should be objects of investigation in the future. 

% correlation energy and performance
Figure \ref{fig:combined} illustrates that the impact of compilation is similar for both NUC and server platforms, and the results regarding execution time correspond closely with energy usage. The linear correlation between energy consumption and execution time indicates that any enhancements in execution time will also improve energy efficiency. We calculated Pearson’s correlation coefficient \cite{wohlinexperimentation} for energy and execution time data for each subject, revealing a strong correlation between the two. While this correlation is helpful, it is not always typical. In scenarios involving dynamically changing frequency, parallelization, and memory-bound workloads, energy consumption can exhibit different trends compared to execution time \cite{jin2017survey}. Therefore, the relationship between these two factors warrants further investigation.

\begin{tcolorbox}%[
%  colback=customblue!10,
%  colframe=customblue!80!black,
%  coltitle=white,
%  colbacktitle=customblue!80!black,
%  fonttitle=\bfseries,
%  left=1mm,
%  right=1mm,
%  top=1mm,
%  bottom=1mm,
%  title=Summary
%]
\textbf{\textit{Summary}} - Compilation can significantly speed up execution compared to interpreted Python code, with improvements varying by benchmark. Codon leads with an average improvement of 94.18\%, followed by PyPy at 86.67\% and Numba at 85.86\%. Energy usage and execution time are strongly correlated.
\end{tcolorbox}

\subsection{Considerations on Memory Usage}

\begin{figure}[h]
    \centering
    \includegraphics[width=\columnwidth]{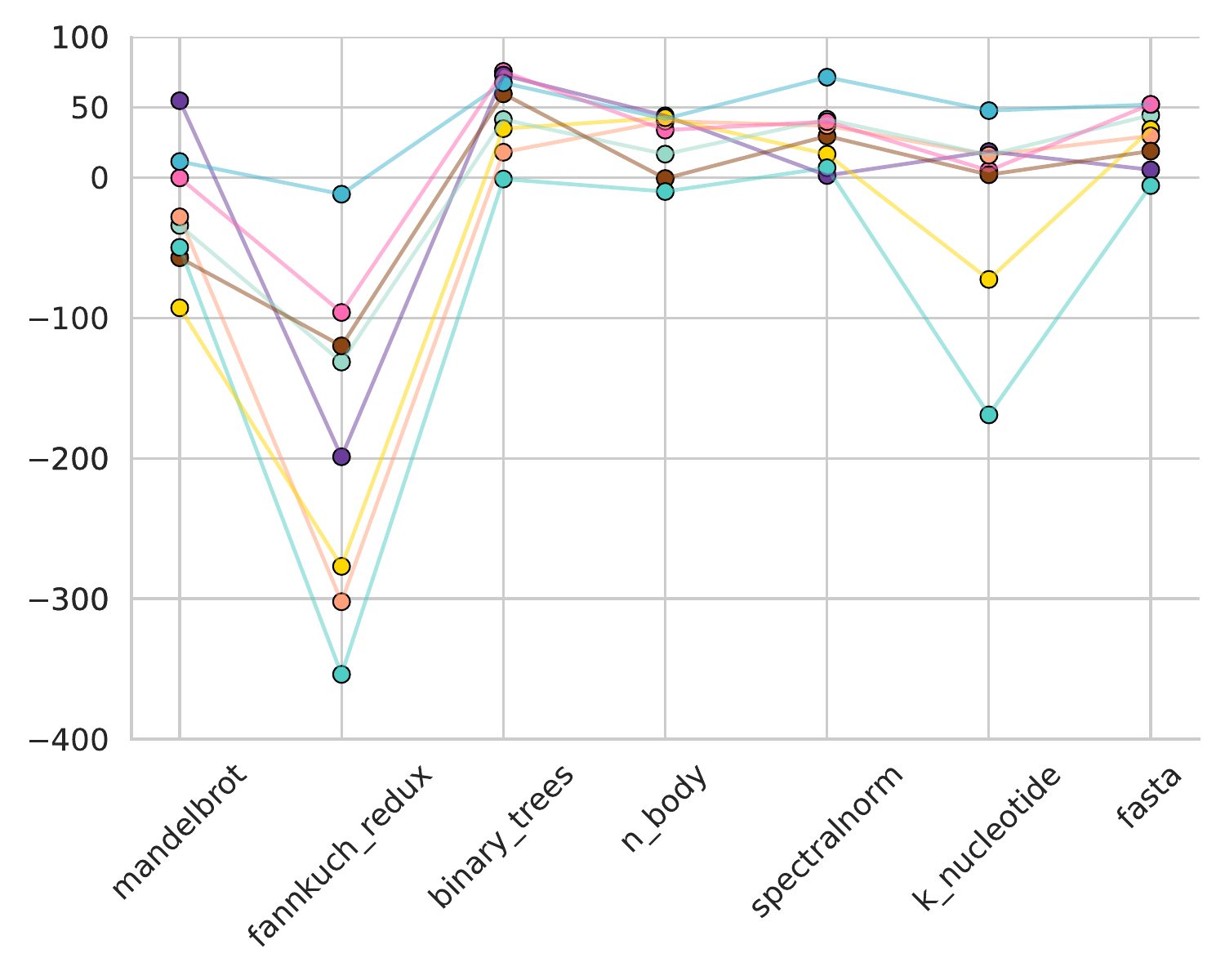}
    \caption{Memory usage improvement on the server across benchmarks compared to CPython.
        Legend: 
        %     ['cpython', 'pypy', 'numba', 'pyston-lite', 'py3.13-jit', 'nuitka', 'cython', 'codon', 'mypyc']
        % \protect\tikz\protect\draw[fill={rgb,255:red,255;green,107;blue,107}] (0,0) rectangle (0.5em,0.5em); CPython,
        \protect\tikz\protect\draw[fill={rgb,255:red,78;green,205;blue,196}] (0,0) rectangle (0.5em,0.5em); PyPy,
        \protect\tikz\protect\draw[fill={rgb,255:red,255;green,160;blue,122}] (0,0) rectangle (0.5em,0.5em); Numba,
        \protect\tikz\protect\draw[fill={rgb,255:red,106;green,61;blue,154}] (0,0) rectangle (0.5em,0.5em); Pyston-lite,
        \protect\tikz\protect\draw[fill={rgb,255:red,255;green,215;blue,0}] (0,0) rectangle (0.5em,0.5em); Python 3.13 JIT,
        \protect\tikz\protect\draw[fill={rgb,255:red,69;green,183;blue,209}] (0,0) rectangle (0.5em,0.5em); Nuitka,
        \protect\tikz\protect\draw[fill={rgb,255:red,152;green,216;blue,200}] (0,0) rectangle (0.5em,0.5em); Cython,
        \protect\tikz\protect\draw[fill={rgb,255:red,255;green,105;blue,180}] (0,0) rectangle (0.5em,0.5em); Codon,
        \protect\tikz\protect\draw[fill={rgb,255:red,139;green,69;blue,19}] (0,0) rectangle (0.5em,0.5em); MyPyc   
    }
    \label{fig:memory_improvement}
    \vspace{-0.5cm}
\end{figure}

The data sorted by subject shows that Nuitka has the lowest average memory usage compared to other compilers, confirmed by Cliff's Delta test indicating a large effect size on the server and a medium effect size on the NUC. Codon also shows a large effect size on the server, while other compilers exhibit small or negligible effects. Individual benchmarks reveal variability in memory usage, some, like fannkuch\_redux and PyPy, exhibit significant increases in memory consumption. These cases influenced the effect size analysis. Figure \ref{fig:memory_improvement} illustrates that all compilers can achieve considerable memory reduction compared to CPython.

It worth noticing that our samples present high variability across repetitions of the same treatment (\ie same compiler and benchmark). In addition, the low memory consumption used by our benchmark (measured in megabytes) makes our results susceptible to unconsidered factors and randomness. In our case, the RSS remains constant with a run but varies across repetitions of the same run (\ie same benchmark and compiler). This is expected due to factors such as memory allocation strategies of Python, garbage collection, and caching strategy during different executions. Python may request memory in larger chunks to the OS, while the garbage collection can be more frequent for certain executions. 

% Despite noticing improvements in memory usage with these three compilers, \textit{we remain uncertain about declaring a positive impact of compilation on memory usage, even though Nuitka consistently improves memory usage across benchmarks and compilers}. The high variability of the memory usage across repetition of the same treatment (\ie same compiler and benchmark) can randomly favor a compiler. Therefore, the mean cannot be used as a reliable statistic to quantify the difference among the groups. As shown in Figure \ref{fig:mem_box}, the data is overlapping and medians are significantly skewed.

%The low memory consumption used by our benchmark (measured in megabytes) makes our results susceptible to unconsidered factors. In our case, the RSS remains constant with a run but varies across repetitions of the same run (\ie same benchmark and compiler). This is expected due to factors such as memory allocation strategies of Python, garbage collection, and caching strategy during different executions. Python may request memory in larger chunks to the OS, while the garbage collection can be more frequent for certain executions. 

We encourage researchers to investigate the impact of compilation on Python code by using benchmarks specifically designed for memory, including those that consume several gigabytes of RSS. It can beneficial to analyze memory allocation patterns, such as the number and size of memory allocation requests (both on the stack and heap), memory fragmentation, paging, and caching. Compilers can affect memory locality and reduce stack usage through techniques such as inlining, loop unrolling, and constant folding.

\begin{tcolorbox}%[
%  colback=customblue!10,
%  colframe=customblue!80!black,
%  coltitle=white,
%  colbacktitle=customblue!80!black,
%  fonttitle=\bfseries,
%  left=1mm,
%  right=1mm,
%  top=1mm,
%  bottom=1mm,
% title=Summary
%]
\textbf{\textit{Summary}} - Memory usage can be significantly improved by compilation. Nuitka achieves over 40\% improvement across the majority of benchmarks. Due to the low memory used by our benchmark, we suggest a more in-depth analysis of memory usage, particularly by analyzing memory-intensive code executions and memory allocation patterns.
\end{tcolorbox}

\subsection{Considerations on LCC miss percentage}
\begin{figure}[ht]
    \centering
    \includegraphics[width=\columnwidth]{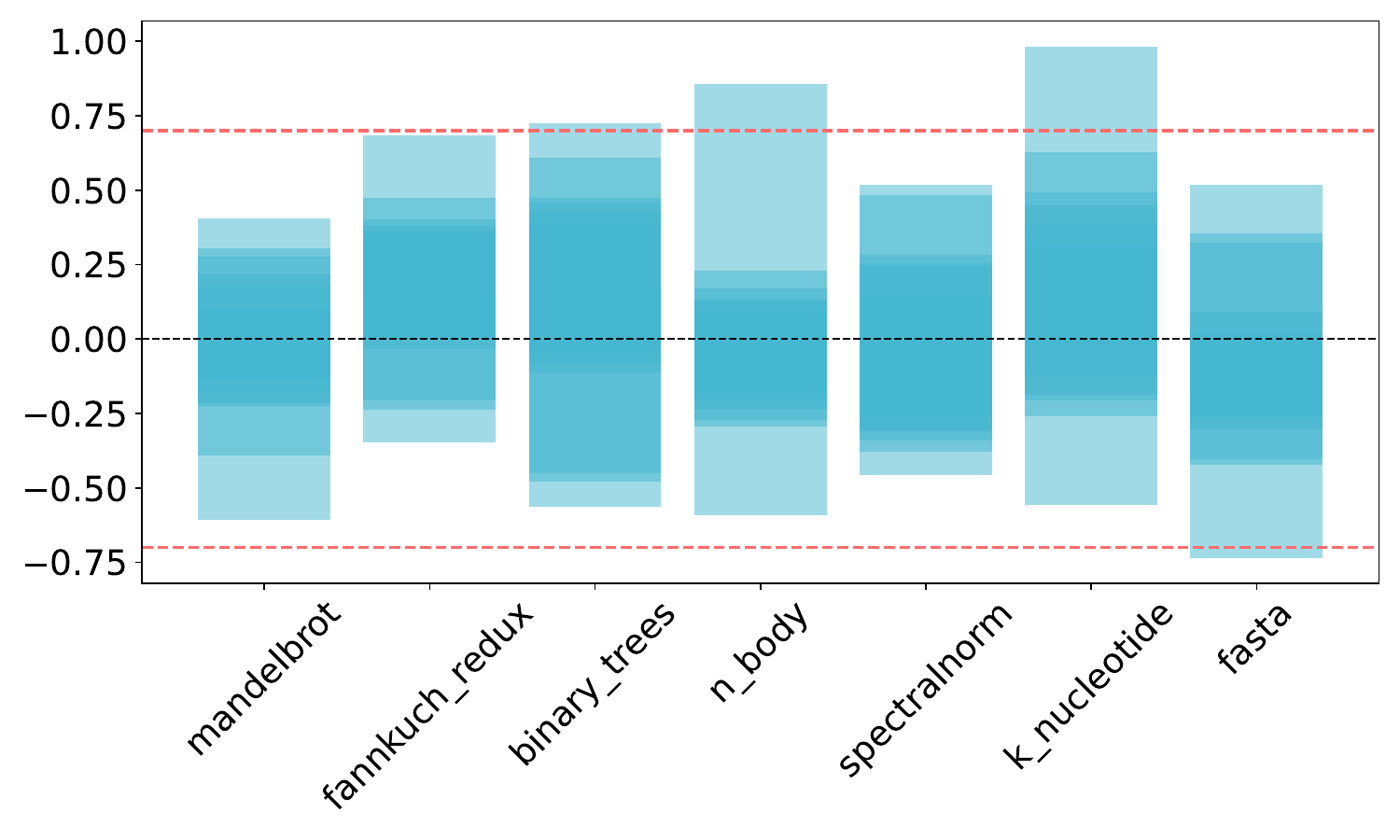}
    \caption{Correlation between execution time and Last-Level Cache misses. Darker areas show higher frequency, with the red dashed line indicating a strong correlation.}
    \vspace{-1cm}
    \label{fig:cache_correlation}
\end{figure}

We cannot recognize any pattern when analyzing the results for LLC miss rate across benchmarks and testbeds. Despite a lower average LLC miss rate for some compilers, such as PyPy, Numba, and Python 3.13 JIT compiler on the server, this result is inconsistent across benchmarks. This result is not present on the NUC, where the aforementioned compilers show a comparable LLC miss rate to CPython. However, it is evident that Nuitka increases LLC miss rate on both testbeds. Therefore, we cannot firmly state that compilation impacts the LLC miss rate in our setting.  The Cliff's Delta effect size yields a small and negligible effect size for all the compilers, except Nuitka where is large on the server. Instead, on the NUC compilation leads to an increase of the LLC miss rate for PyPy, Nuitka, Codon, and Numba.

We analyzed the LCC miss rate to assess how compilers affect this metric and their potential influence on execution time. Figure \ref{fig:cache_correlation} shows Pearson's correlation between LLC miss rate and execution time across benchmarks. We calculated the correlation separately for each testbed and then aggregated and plotted the results. A darker color represents a higher frequency of values. We notice a higher frequency of positive correlation values, meaning a higher rate of LLC miss increases execution time. However, most values are concentrated in ranges suggesting small and moderate positive correlations.  Our results include several small negative correlations that make it infeasible to infer something. Additionally, correlation does not mean causation, so we suggest further investigation of the impact of LLC misses rate on Python code performance in the future. Overall, the trend shows that the metrics could be related, and this is strengthened by Figure \ref{fig:gl2_energy}, which indicates that uncore components, including the cache, consume a big part of the energy.

% Using CPython, developers can use the decorators \texttt{@cache} and \texttt{@lru\_cache} from the \texttt{functools} library to optimize caching. 

\begin{tcolorbox}%[
%  colback=customblue!10,
%  colframe=customblue!80!black,
%  coltitle=white,
%  colbacktitle=customblue!80!black,
%  fonttitle=\bfseries,
%  left=1mm,
%  right=1mm,
%  top=1mm,
%  bottom=1mm,
%  title=Summary
%]
\textbf{\textit{Summary}} - The impact of compilation on the Last-Level Cache (LLC) miss rate is unclear due to inconsistent results across benchmarks and testbeds. There is a positive correlation between LLC miss rate, execution time, and energy usage, with most energy consumed by uncore components like the LLC. Further investigation is recommended on how caching affects energy usage and execution time.
\end{tcolorbox}

\subsection{Implications for Python Practitioners}
Our results provide some practical insights for Python practitioners (including researchers). We suggest using PyPy, Numba, and Codon, as they significantly improve Python efficiency. These three compilers present a different learning curve.
% In this study, we used code that can be compiled without changing its behavior, such as the control flow, not to introduce accidental optimization that could bias our results. 
Based on our experience, Numba and Codon need more profound knowledge of their features and considerable code modifications. With Numba and Codon, the developers need to include specific decorators, such as \texttt{@numba.jit} and \texttt{@codon.jit}, before the definition of the function to optimize. Numba and Codon support adopting unboxed types (\eg int32) that developers can be manually specified. Additionally, Codon can have compatibility problems with CPython as it is built on a different implementation of Python, and, at its current state, it has limited optimization with popular libraries, such as NumPy. Numba and Codon can be considered strong choices for optimizing Python code, especially for compute-intensive code, and in our experiment, outperform PyPy. Conversely, PyPy combines compatibility with CPython, despite being restricted to RPython, minimal code changes, and efficiency improvements. The significant improvements achieved by using Codon and PyPy also indicate that using an alternative implementation of Python may be a good idea to optimize performance and energy usage. 

%Our study includes four JIT compilers and four AOT compilers. When considering energy consumption and execution time, the type of compiler appears to have minimal impact, as the top three compilers are two JIT, Numba, and Pypy, along with the AOT compiler Codon. However, for memory usage, AOT compilers seem to perform slightly better, with the leading three compilers being AOT, Codon, Mypyc, and Cython. We recommend that researchers conduct further investigations into how the type of compilation affects the memory usage of Python code. Specifically, AOT compilers facilitate static type checking, which can help conserve memory.

According to our results, code characteristics play a significant role in defining the magnitude of the compiler impact. Additionally, there are cases where the compilers introduce inefficiencies. For example, the experimental JIT compiler integrated with Python 3.13 provides an average energy improvement of over 40\% for fannkuch\_redux and binary\_trees. However, it introduces overhead and increases energy usage for mandelbrot, k\_knucleotide, and fasta. In addition, its impact on memory usage is small or negative. We recommend that researchers and developers explore how code characteristics relate to compilation methods in order to maximize the benefits obtained from them.
\section{Threats to Validity}
\label{ss:ttv}
We discuss potential threats to the validity of this study and how we mitigate them. We follow the classification provided by Cook and Campbell \cite{Cook:1979} and elaborated by Wohlin \etal \cite{wohlinexperimentation}.

\subsection{Internal Validity}
We adopted the code of benchmarks to be compatible with Numba and Codon. The modifications involved data types, classes, and printing, which were adapted due to the characteristics of the compilers. While adjusting the code for these compilers, there may have been unintended inefficiencies or performance improvements introduced during the process. To prevent any erroneous behavior, we tested the functions and kept changes to a minimum, ensuring that we did not alter the control and data flow of the code. The data comparison between Pyston Lite and CPython is surprising, as we expected Pyston Lite to perform better. This result might be affected by an unchecked factor, but we confirmed that Pyston Lite was disabled during the CPython test. The data on memory usage and last-level cache (LLC) misses varies widely across benchmarks and testbeds due to factors like memory management by the operating system and Python, which were uncontrolled in our experiment. We used EnergiBridge for memory profiling to track both physical and virtual memory usage. Due to significant variability, combining EnergiBridge output with additional metrics, such as paging, would have helped reveal data patterns.

\subsection{External Validity}
Our results cannot be generalized to every Python compiler and code characteristics, as we only analyzed a small sample of eight compilers and seven benchmarks. However, our selection includes many of the most widely used and diverse Python compilers. We use code commonly found in scientific domains and in studies of software performance and energy efficiency \cite{pereira2021ranking}. The benchmarks address non-trivial problems that reflect real-world computational challenges. Despite the different hardware architecture, the experiment must be generalized on more than two testbeds. The results show consistent energy and execution time on both testbeds. Although the testbeds have different CPUs, this outcome can be due to the similar memory size of the testbeds and the constrained number of cores.

\subsection{Construct Validity}
The characteristics of the code may have influenced our experimental results, favoring compilers that can better improve code with specific characteristics. We chose the benchmarks based on their high computational demand to highlight possible optimizations more effectively. Furthermore, some Python compilers show greater improvements when utilizing third-party functions and parallelism, as they are often employed in high-performance computing environments. For instance, Numba is well-known for its ability to optimize NumPy operations and support parallel computing. We excluded these characteristics as they can influence Python code efficiency.

\subsection{Conclusion Validity}
Our experimental results are significantly influenced by the reliability of our measurement tools and the sample size. The measurement tools we use, namely, EnergiBridge, perf, and time, are known for their reliability and are commonly employed in this type of study. We perform each combination of benchmark and compiler on the server for 10 iterations and on the NUC for 15 iterations. Nevertheless, this number of trials might not be adequate to reveal a definitive trend in the results. The results for execution time and energy usage are net, allowing us to uncover a pattern. 
\section{Conclusion and Future Work}
\label{s:conclusion}
Research on Python highlights its energy inefficiency and performance issues. While Python is vital for automation in areas like scientific software and machine learning, the benefits of compilation on its energy usage remain unquantified. Existing studies often overlook variables affecting performance, such as the number of active cores and CPU frequency. We compared eight Python compilers based on execution time, energy consumption, memory usage, and Last-Level Cache (LLC) miss rate across seven benchmarks, using CPython as a control. Our findings indicate that compilation significantly enhances performance and energy efficiency, with PyPy, Codon, and Numba showing over 90\% improvement on the majority of benchmarks, while Nuitka consistently improves memory usage across testbeds for the majority of benchmarks. However, LLC miss rate results were inconsistent across benchmarks and testbeds.

The relationship between code characteristics and the approaches used by various compilers should be more thoroughly investigated in the future as the results are affected. In addition, our results show that uncore components, such as LLC, may play a primary role in optimizing the energy use and performance of Python code. A future experiment can use a set of benchmarks picked specifically to stress uncore components. Another factor that requires further investigation is its relationship with compilers and platform characteristics, as our results from the NUC differ significantly from those of the server.

\section*{Acknowledgement}
This project has received funding from the LETSGO Project promoted by the Netherlands Enterprise Agency (Rijksdienst voor Ondernemend Nederland).

\bibliographystyle{ACM-Reference-Format}
\bibliography{refs}


\begin{thebibliography}{54}

%%% the Web address.

\ifx \showCODEN    \undefined \def \showCODEN     #1{\unskip}     \fi
\ifx \showDOI      \undefined \def \showDOI       #1{#1}\fi
\ifx \showISBNx    \undefined \def \showISBNx     #1{\unskip}     \fi
\ifx \showISBNxiii \undefined \def \showISBNxiii  #1{\unskip}     \fi
\ifx \showISSN     \undefined \def \showISSN      #1{\unskip}     \fi
\ifx \showLCCN     \undefined \def \showLCCN      #1{\unskip}     \fi
\ifx \shownote     \undefined \def \shownote      #1{#1}          \fi
\ifx \showarticletitle \undefined \def \showarticletitle #1{#1}   \fi
\ifx \showURL      \undefined \def \showURL       {\relax}        \fi
% invisible to TeX
\providecommand\bibfield[2]{#2}
\providecommand\bibinfo[2]{#2}
\providecommand\natexlab[1]{#1}
\providecommand\showeprint[2][]{arXiv:#2}

\bibitem[Abdulsalam et~al\mbox{.}(2014)]%
        {abdulsalam2014program}
\bibfield{author}{\bibinfo{person}{Sarah Abdulsalam}, \bibinfo{person}{Donna Lakomski}, \bibinfo{person}{Qijun Gu}, \bibinfo{person}{Tongdan Jin}, {and} \bibinfo{person}{Ziliang Zong}.} \bibinfo{year}{2014}\natexlab{}.
\newblock \showarticletitle{Program energy efficiency: The impact of language, compiler and implementation choices}. In \bibinfo{booktitle}{\emph{International Green Computing Conference}}. IEEE, \bibinfo{pages}{1--6}.
\newblock


\bibitem[Akeret et~al\mbox{.}(2015)]%
        {akeret2015hope}
\bibfield{author}{\bibinfo{person}{Jo{\"e}l Akeret}, \bibinfo{person}{Lukas Gamper}, \bibinfo{person}{Adam Amara}, {and} \bibinfo{person}{Alexandre Refregier}.} \bibinfo{year}{2015}\natexlab{}.
\newblock \showarticletitle{HOPE: A Python just-in-time compiler for astrophysical computations}.
\newblock \bibinfo{journal}{\emph{Astronomy and Computing}}  \bibinfo{volume}{10} (\bibinfo{year}{2015}), \bibinfo{pages}{1--8}.
\newblock


\bibitem[Augier et~al\mbox{.}(2021)]%
        {augier2021reducing}
\bibfield{author}{\bibinfo{person}{Pierre Augier}, \bibinfo{person}{Carl~Friedrich Bolz-Tereick}, \bibinfo{person}{Serge Guelton}, {and} \bibinfo{person}{Ashwin~Vishnu Mohanan}.} \bibinfo{year}{2021}\natexlab{}.
\newblock \showarticletitle{Reducing the ecological impact of computing through education and Python compilers}.
\newblock \bibinfo{journal}{\emph{Nature Astronomy}} \bibinfo{volume}{5}, \bibinfo{number}{4} (\bibinfo{year}{2021}), \bibinfo{pages}{334--335}.
\newblock


\bibitem[Behnel et~al\mbox{.}(2010)]%
        {behnel2010cython}
\bibfield{author}{\bibinfo{person}{Stefan Behnel}, \bibinfo{person}{Robert Bradshaw}, \bibinfo{person}{Craig Citro}, \bibinfo{person}{Lisandro Dalcin}, \bibinfo{person}{Dag~Sverre Seljebotn}, {and} \bibinfo{person}{Kurt Smith}.} \bibinfo{year}{2010}\natexlab{}.
\newblock \showarticletitle{Cython: The best of both worlds}.
\newblock \bibinfo{journal}{\emph{Computing in Science \& Engineering}} \bibinfo{volume}{13}, \bibinfo{number}{2} (\bibinfo{year}{2010}), \bibinfo{pages}{31--39}.
\newblock


\bibitem[Castor(2024)]%
        {castor2024estimating}
\bibfield{author}{\bibinfo{person}{Fernando Castor}.} \bibinfo{year}{2024}\natexlab{}.
\newblock \showarticletitle{Estimating the Energy Footprint of Software Systems: a Primer}.
\newblock \bibinfo{journal}{\emph{arXiv preprint arXiv:2407.11611}} (\bibinfo{year}{2024}).
\newblock


\bibitem[Cook and Campbell(1979)]%
        {Cook:1979}
\bibfield{author}{\bibinfo{person}{{Thomas D} Cook} {and} \bibinfo{person}{{D T} Campbell}.} \bibinfo{year}{1979}\natexlab{}.
\newblock \bibinfo{booktitle}{\emph{Quasi-Experimentation: Design and Analysis Issues for Field Settings}}.
\newblock \bibinfo{publisher}{Houghton Mifflin}.
\newblock


\bibitem[Courtines et~al\mbox{.}(2021)]%
        {PLCB}
\bibfield{author}{\bibinfo{person}{Elana Courtines}, \bibinfo{person}{Georges Da~Costa}, {and} \bibinfo{person}{Patrica Stolf}.} \bibinfo{year}{2021}\natexlab{}.
\newblock \bibinfo{title}{Programming Language and Compiler Benchmarks}.
\newblock \bibinfo{howpublished}{{https://hal.science/hal-04610856v1/document}}.
\newblock
\newblock
\shownote{Accessed: 2025-01-1}.


\bibitem[Cruz(2021)]%
        {cruz2021green}
\bibfield{author}{\bibinfo{person}{Lu{\'\i}s Cruz}.} \bibinfo{year}{2021}\natexlab{}.
\newblock \showarticletitle{Green software engineering done right: a scientific guide to set up energy efficiency experiments}.
\newblock \bibinfo{journal}{\emph{Blog post}} (\bibinfo{year}{2021}).
\newblock


\bibitem[Drepper(2007)]%
        {drepper2007every}
\bibfield{author}{\bibinfo{person}{Ulrich Drepper}.} \bibinfo{year}{2007}\natexlab{}.
\newblock \showarticletitle{What every programmer should know about memory}.
\newblock \bibinfo{journal}{\emph{Red Hat, Inc}} \bibinfo{volume}{11}, \bibinfo{number}{2007} (\bibinfo{year}{2007}), \bibinfo{pages}{2007}.
\newblock


\bibitem[Durieux(2024)]%
        {energibridge}
\bibfield{author}{\bibinfo{person}{Thomas Durieux}.} \bibinfo{year}{2024}\natexlab{}.
\newblock \bibinfo{booktitle}{\emph{EnergiBridge Measurement Utility}}.
\newblock
\urldef\tempurl%
\url{https://github.com/tdurieux/EnergiBridge}
\showURL{%
\tempurl}


\bibitem[Forcier(2024)]%
        {paramiko}
\bibfield{author}{\bibinfo{person}{Jeff Forcier}.} \bibinfo{year}{2024}\natexlab{}.
\newblock \bibinfo{booktitle}{\emph{Paramiko Python Package}}.
\newblock
\urldef\tempurl%
\url{https://github.com/paramiko/paramiko}
\showURL{%
\tempurl}


\bibitem[Foundation(2025)]%
        {pyperformance}
\bibfield{author}{\bibinfo{person}{Python~Software Foundation}.} \bibinfo{year}{2025}\natexlab{}.
\newblock \bibinfo{title}{PyPerformance: Python Performance Benchmark Suite}.
\newblock \bibinfo{howpublished}{\url{https://github.com/python/pyperformance}}.
\newblock
\newblock
\shownote{Accessed: 2025-01-13}.


\bibitem[Georgiou et~al\mbox{.}(2017)]%
        {georgiou2017analyzing}
\bibfield{author}{\bibinfo{person}{Stefanos Georgiou}, \bibinfo{person}{Maria Kechagia}, {and} \bibinfo{person}{Diomidis Spinellis}.} \bibinfo{year}{2017}\natexlab{}.
\newblock \showarticletitle{Analyzing programming languages' energy consumption: An empirical study}. In \bibinfo{booktitle}{\emph{Proceedings of the 21st Pan-Hellenic Conference on Informatics}}. \bibinfo{pages}{1--6}.
\newblock


\bibitem[Gouy({[n.\,d.]})]%
        {CLBG}
\bibfield{author}{\bibinfo{person}{Isaac Gouy}.} \bibinfo{year}{[n.\,d.]}\natexlab{}.
\newblock \bibinfo{title}{The Computer Language Benchmarks Game}.
\newblock \bibinfo{howpublished}{{https://benchmarksgame-team.pages.debian.net/benchmarksgame/}}.
\newblock
\newblock
\shownote{Accessed: 2025-01-1}.


\bibitem[Gregg(2019)]%
        {gregg2019bpf}
\bibfield{author}{\bibinfo{person}{Brendan Gregg}.} \bibinfo{year}{2019}\natexlab{}.
\newblock \bibinfo{booktitle}{\emph{BPF performance tools}}.
\newblock \bibinfo{publisher}{Addison-Wesley Professional}.
\newblock


\bibitem[Guelton et~al\mbox{.}(2015)]%
        {Guelton2015Pythran}
\bibfield{author}{\bibinfo{person}{Serge Guelton}, \bibinfo{person}{Pierrick Brunet}, \bibinfo{person}{Mehdi Amini}, \bibinfo{person}{Adrien Merlini}, \bibinfo{person}{Xavier Corbillon}, {and} \bibinfo{person}{Alan Raynaud}.} \bibinfo{year}{2015}\natexlab{}.
\newblock \showarticletitle{Pythran: Enabling static optimization of scientific Python programs}.
\newblock \bibinfo{journal}{\emph{Computational Science \& Discovery}} \bibinfo{volume}{8}, \bibinfo{number}{1} (\bibinfo{year}{2015}), \bibinfo{pages}{014001}.
\newblock
\urldef\tempurl%
\url{https://doi.org/10.1088/1749-4680/8/1/014001}
\showDOI{\tempurl}


\bibitem[Ibrahim et~al\mbox{.}(2016)]%
        {ibrahim2016governing}
\bibfield{author}{\bibinfo{person}{Shadi Ibrahim}, \bibinfo{person}{Tien-Dat Phan}, \bibinfo{person}{Alexandra Carpen-Amarie}, \bibinfo{person}{Houssem-Eddine Chihoub}, \bibinfo{person}{Diana Moise}, {and} \bibinfo{person}{Gabriel Antoniu}.} \bibinfo{year}{2016}\natexlab{}.
\newblock \showarticletitle{Governing energy consumption in Hadoop through CPU frequency scaling: An analysis}.
\newblock \bibinfo{journal}{\emph{Future Generation Computer Systems}}  \bibinfo{volume}{54} (\bibinfo{year}{2016}), \bibinfo{pages}{219--232}.
\newblock


\bibitem[{IronPython Community}(2024)]%
        {ironpython}
\bibfield{author}{\bibinfo{person}{{IronPython Community}}.} \bibinfo{year}{2024}\natexlab{}.
\newblock \bibinfo{title}{{IronPython}: An implementation of {Python} for the .{NET} Framework}.
\newblock \bibinfo{howpublished}{\url{https://ironpython.net/}}.
\newblock
\newblock
\shownote{Accessed: 2024-01-14}.


\bibitem[Jin et~al\mbox{.}(2017)]%
        {jin2017survey}
\bibfield{author}{\bibinfo{person}{Chao Jin}, \bibinfo{person}{Bronis~R de Supinski}, \bibinfo{person}{David Abramson}, \bibinfo{person}{Heidi Poxon}, \bibinfo{person}{Luiz DeRose}, \bibinfo{person}{Minh~Ngoc Dinh}, \bibinfo{person}{Mark Endrei}, {and} \bibinfo{person}{Elizabeth~R Jessup}.} \bibinfo{year}{2017}\natexlab{}.
\newblock \showarticletitle{A survey on software methods to improve the energy efficiency of parallel computing}.
\newblock \bibinfo{journal}{\emph{The International Journal of High Performance Computing Applications}} \bibinfo{volume}{31}, \bibinfo{number}{6} (\bibinfo{year}{2017}), \bibinfo{pages}{517--549}.
\newblock


\bibitem[Juneau et~al\mbox{.}(2010)]%
        {juneau2010definitive}
\bibfield{author}{\bibinfo{person}{Josh Juneau}, \bibinfo{person}{Jim Baker}, \bibinfo{person}{Frank Wierzbicki}, \bibinfo{person}{Leo Soto}, {and} \bibinfo{person}{Victor Ng}.} \bibinfo{year}{2010}\natexlab{}.
\newblock \bibinfo{booktitle}{\emph{The Definitive Guide to {Jython}: {Python} for the {Java} Platform}}.
\newblock \bibinfo{publisher}{Apress}.
\newblock
\showISBNx{978-1-4302-2527-0}


\bibitem[Koedijk and Oprescu(2022)]%
        {koedijk2022finding}
\bibfield{author}{\bibinfo{person}{Lukas Koedijk} {and} \bibinfo{person}{Ana Oprescu}.} \bibinfo{year}{2022}\natexlab{}.
\newblock \showarticletitle{Finding significant differences in the energy consumption when comparing programming languages and programs}. In \bibinfo{booktitle}{\emph{2022 International Conference on ICT for Sustainability (ICT4S)}}. IEEE, \bibinfo{pages}{1--12}.
\newblock


\bibitem[Lam et~al\mbox{.}(2015)]%
        {lam2015numba}
\bibfield{author}{\bibinfo{person}{Siu~Kwan Lam}, \bibinfo{person}{Antoine Pitrou}, {and} \bibinfo{person}{Stanley Seibert}.} \bibinfo{year}{2015}\natexlab{}.
\newblock \showarticletitle{Numba: A llvm-based python jit compiler}. In \bibinfo{booktitle}{\emph{Proceedings of the Second Workshop on the LLVM Compiler Infrastructure in HPC}}. \bibinfo{pages}{1--6}.
\newblock


\bibitem[Malavolta et~al\mbox{.}(2024)]%
        {malavolta2024ten}
\bibfield{author}{\bibinfo{person}{Ivano Malavolta}, \bibinfo{person}{Vincenzo Stoico}, {and} \bibinfo{person}{Patricia Lago}.} \bibinfo{year}{2024}\natexlab{}.
\newblock \showarticletitle{Ten Years of Teaching Empirical Software Engineering in the Context of Energy-Efficient Software}.
\newblock In \bibinfo{booktitle}{\emph{Handbook on Teaching Empirical Software Engineering}}. \bibinfo{publisher}{Springer}, \bibinfo{pages}{209--253}.
\newblock


\bibitem[Melan{\c{c}}on et~al\mbox{.}(2023)]%
        {melanccon2023executable}
\bibfield{author}{\bibinfo{person}{Olivier Melan{\c{c}}on}, \bibinfo{person}{Marc Feeley}, {and} \bibinfo{person}{Manuel Serrano}.} \bibinfo{year}{2023}\natexlab{}.
\newblock \showarticletitle{An Executable Semantics for Faster Development of Optimizing Python Compilers}. In \bibinfo{booktitle}{\emph{Proceedings of the 16th ACM SIGPLAN International Conference on Software Language Engineering}}. \bibinfo{pages}{15--28}.
\newblock


\bibitem[Merelo-Guerv{\'o}s et~al\mbox{.}(2016)]%
        {merelo2016ranking}
\bibfield{author}{\bibinfo{person}{Juan~Juli{\'a}n Merelo-Guerv{\'o}s}, \bibinfo{person}{Israel Blancas-Alvarez}, \bibinfo{person}{Pedro~A Castillo}, \bibinfo{person}{Gustavo Romero}, \bibinfo{person}{Pablo Garc{\'\i}a-S{\'a}nchez}, \bibinfo{person}{Victor~M Rivas}, \bibinfo{person}{Mario Garc{\'\i}a-Valdez}, \bibinfo{person}{Amaury Hern{\'a}ndez-{\'A}guila}, {and} \bibinfo{person}{Mario Rom{\'a}n}.} \bibinfo{year}{2016}\natexlab{}.
\newblock \showarticletitle{Ranking the Performance of Compiled and Interpreted Languages in Genetic Algorithms}. In \bibinfo{booktitle}{\emph{Proceedings of the International Conference on Evolutionary Computation Theory and Applications, Porto, Portugal}}, Vol.~\bibinfo{volume}{11}. \bibinfo{pages}{164--170}.
\newblock


\bibitem[Molnár(2024)]%
        {perf}
\bibfield{author}{\bibinfo{person}{Ingo Molnár}.} \bibinfo{year}{2024}\natexlab{}.
\newblock \bibinfo{booktitle}{\emph{Performance analysis tools for Linux}}.
\newblock
\urldef\tempurl%
\url{https://man7.org/linux/man-pages/man1/perf.1.html}
\showURL{%
\tempurl}


\bibitem[{mypy developers}(2025)]%
        {mypyc}
\bibfield{author}{\bibinfo{person}{{mypy developers}}.} \bibinfo{year}{2025}\natexlab{}.
\newblock \bibinfo{booktitle}{\emph{mypyc: Compile Python Modules to C Extensions}}.
\newblock
\urldef\tempurl%
\url{https://github.com/mypyc/mypyc}
\showURL{%
\tempurl}


\bibitem[Nanz and Furia(2015)]%
        {nanz2015comparative}
\bibfield{author}{\bibinfo{person}{Sebastian Nanz} {and} \bibinfo{person}{Carlo~A Furia}.} \bibinfo{year}{2015}\natexlab{}.
\newblock \showarticletitle{A comparative study of programming languages in rosetta code}. In \bibinfo{booktitle}{\emph{2015 IEEE/ACM 37th IEEE International Conference on Software Engineering}}, Vol.~\bibinfo{volume}{1}. IEEE, \bibinfo{pages}{778--788}.
\newblock


\bibitem[{Nuitka Developers}(2024)]%
        {Nuitka2024}
\bibfield{author}{\bibinfo{person}{{Nuitka Developers}}.} \bibinfo{year}{2024}\natexlab{}.
\newblock \bibinfo{title}{Nuitka - The Python Compiler}.
\newblock \bibinfo{howpublished}{\url{https://nuitka.net/}}.
\newblock
\newblock
\shownote{[Accessed: 2024-03-05]}.


\bibitem[{Oracle}(2024)]%
        {graalpy}
\bibfield{author}{\bibinfo{person}{{Oracle}}.} \bibinfo{year}{2024}\natexlab{}.
\newblock \bibinfo{title}{{GraalPy}: A high-performance {Python} implementation for the {JVM}}.
\newblock \bibinfo{howpublished}{\url{https://www.graalvm.org/python/}}.
\newblock
\newblock
\shownote{Accessed: 2024-01-14}.


\bibitem[package(2025)]%
        {repl-package}
\bibfield{author}{\bibinfo{person}{Replication package}.} \bibinfo{year}{2025}\natexlab{}.
\newblock \bibinfo{title}{Replication package for this study}.
\newblock
\newblock
\urldef\tempurl%
\url{https://github.com/S2-group/python-compilers-rep-pkg}
\showURL{%
\tempurl}


\bibitem[Peng et~al\mbox{.}(2024)]%
        {peng2024less}
\bibfield{author}{\bibinfo{person}{Yun Peng}, \bibinfo{person}{Ruida Hu}, \bibinfo{person}{Ruoke Wang}, \bibinfo{person}{Cuiyun Gao}, \bibinfo{person}{Shuqing Li}, {and} \bibinfo{person}{Michael~R Lyu}.} \bibinfo{year}{2024}\natexlab{}.
\newblock \showarticletitle{Less is More? An Empirical Study on Configuration Issues in Python PyPI Ecosystem}. In \bibinfo{booktitle}{\emph{Proceedings of the IEEE/ACM 46th International Conference on Software Engineering}}. \bibinfo{pages}{1--12}.
\newblock


\bibitem[Pereira et~al\mbox{.}(2021)]%
        {pereira2021ranking}
\bibfield{author}{\bibinfo{person}{Rui Pereira}, \bibinfo{person}{Marco Couto}, \bibinfo{person}{Francisco Ribeiro}, \bibinfo{person}{Rui Rua}, \bibinfo{person}{J{\'a}come Cunha}, \bibinfo{person}{Jo{\~a}o~Paulo Fernandes}, {and} \bibinfo{person}{Jo{\~a}o Saraiva}.} \bibinfo{year}{2021}\natexlab{}.
\newblock \showarticletitle{Ranking programming languages by energy efficiency}.
\newblock \bibinfo{journal}{\emph{Science of Computer Programming}}  \bibinfo{volume}{205} (\bibinfo{year}{2021}), \bibinfo{pages}{102609}.
\newblock


\bibitem[Pfeiffer(2024)]%
        {pfeiffer2024energy}
\bibfield{author}{\bibinfo{person}{Rolf-Helge Pfeiffer}.} \bibinfo{year}{2024}\natexlab{}.
\newblock \showarticletitle{On the Energy Consumption of CPython}. In \bibinfo{booktitle}{\emph{International Conference on the Quality of Information and Communications Technology}}. Springer, \bibinfo{pages}{194--209}.
\newblock


\bibitem[Portegies~Zwart(2020)]%
        {portegies2020ecological}
\bibfield{author}{\bibinfo{person}{Simon Portegies~Zwart}.} \bibinfo{year}{2020}\natexlab{}.
\newblock \showarticletitle{The ecological impact of high-performance computing in astrophysics}.
\newblock \bibinfo{journal}{\emph{Nature Astronomy}} \bibinfo{volume}{4}, \bibinfo{number}{9} (\bibinfo{year}{2020}), \bibinfo{pages}{819--822}.
\newblock


\bibitem[Pouyeh(2024)]%
        {pouyeh2024impact}
\bibfield{author}{\bibinfo{person}{Banijamali Pouyeh}.} \bibinfo{year}{2024}\natexlab{}.
\newblock \showarticletitle{On the impact of Codon compilation on energy consumption and performance of Python code}.
\newblock  (\bibinfo{year}{2024}).
\newblock


\bibitem[{PyPy Developers}(2024)]%
        {PyPy2024}
\bibfield{author}{\bibinfo{person}{{PyPy Developers}}.} \bibinfo{year}{2024}\natexlab{}.
\newblock \bibinfo{title}{PyPy - Fast, flexible, and compliant Python interpreter}.
\newblock \bibinfo{howpublished}{\url{https://www.pypy.org/}}.
\newblock
\newblock
\shownote{[Accessed: 2024-03-05]}.


\bibitem[{Pyston developers}(2025)]%
        {pyston-lite}
\bibfield{author}{\bibinfo{person}{{Pyston developers}}.} \bibinfo{year}{2025}\natexlab{}.
\newblock \bibinfo{booktitle}{\emph{Pyston-lite: Python JIT as an Extension Module}}.
\newblock
\urldef\tempurl%
\url{https://github.com/pyston/pyston}
\showURL{%
\tempurl}


\bibitem[{Python Core Developers}(2024)]%
        {python313jit}
\bibfield{author}{\bibinfo{person}{{Python Core Developers}}.} \bibinfo{year}{2024}\natexlab{}.
\newblock \showarticletitle{Experimental {JIT} Compiler in {Python} 3.13}. In \bibinfo{booktitle}{\emph{Python 3.13 Release}}. \bibinfo{publisher}{Python Software Foundation}.
\newblock
\newblock
\shownote{Experimental feature}.


\bibitem[Reya et~al\mbox{.}(2023)]%
        {reya2023greenpy}
\bibfield{author}{\bibinfo{person}{Nurzihan~Fatema Reya}, \bibinfo{person}{Abtahi Ahmed}, \bibinfo{person}{Tashfia Zaman}, {and} \bibinfo{person}{Md~Motaharul Islam}.} \bibinfo{year}{2023}\natexlab{}.
\newblock \showarticletitle{GreenPy: evaluating application-level energy efficiency in Python for green computing}.
\newblock \bibinfo{journal}{\emph{Annals of Emerging Technologies in Computing (AETiC)}} \bibinfo{volume}{7}, \bibinfo{number}{3} (\bibinfo{year}{2023}), \bibinfo{pages}{92--110}.
\newblock


\bibitem[Saedi(2024)]%
        {saeditowards}
\bibfield{author}{\bibinfo{person}{Omid Saedi}.} \bibinfo{year}{2024}\natexlab{}.
\newblock \showarticletitle{Towards Eco-Conscious Python: A Comparative Analysis of Performance, Energy Efficiency and Carbon Emissions Between CPython and Alternative Implementations}.
\newblock  (\bibinfo{year}{2024}).
\newblock


\bibitem[Shajii et~al\mbox{.}(2019)]%
        {shajii2019seq}
\bibfield{author}{\bibinfo{person}{Ariya Shajii}, \bibinfo{person}{Ibrahim Numanagi{\'c}}, \bibinfo{person}{Riyadh Baghdadi}, \bibinfo{person}{Bonnie Berger}, {and} \bibinfo{person}{Saman Amarasinghe}.} \bibinfo{year}{2019}\natexlab{}.
\newblock \showarticletitle{Seq: a high-performance language for bioinformatics}.
\newblock \bibinfo{journal}{\emph{Proceedings of the ACM on programming languages}} \bibinfo{volume}{3}, \bibinfo{number}{OOPSLA} (\bibinfo{year}{2019}), \bibinfo{pages}{1--29}.
\newblock


\bibitem[Shajii et~al\mbox{.}(2023)]%
        {shajii2023codon}
\bibfield{author}{\bibinfo{person}{Ariya Shajii}, \bibinfo{person}{Gabriel Ramirez}, \bibinfo{person}{Haris Smajlovi{\'c}}, \bibinfo{person}{Jessica Ray}, \bibinfo{person}{Bonnie Berger}, \bibinfo{person}{Saman Amarasinghe}, {and} \bibinfo{person}{Ibrahim Numanagi{\'c}}.} \bibinfo{year}{2023}\natexlab{}.
\newblock \showarticletitle{Codon: A compiler for high-performance pythonic applications and dsls}. In \bibinfo{booktitle}{\emph{Proceedings of the 32nd ACM SIGPLAN International Conference on Compiler Construction}}. \bibinfo{pages}{191--202}.
\newblock


\bibitem[Software and Sustainability~Group(2025)]%
        {experiment-runner}
\bibfield{author}{\bibinfo{person}{Software} {and} \bibinfo{person}{Vrije Universiteit~Amsterdam Sustainability~Group}.} \bibinfo{year}{2025}\natexlab{}.
\newblock \bibinfo{booktitle}{\emph{Experiment Runner}}.
\newblock
\urldef\tempurl%
\url{https://github.com/S2-group/experiment-runner}
\showURL{%
\tempurl}


\bibitem[{Stack Overflow}(2024)]%
        {stackoverflow2024}
\bibfield{author}{\bibinfo{person}{{Stack Overflow}}.} \bibinfo{year}{2024}\natexlab{}.
\newblock \bibinfo{title}{Stack Overflow Developer Survey 2024}.
\newblock
\newblock
\urldef\tempurl%
\url{https://survey.stackoverflow.co/2024/}
\showURL{%
\tempurl}
\newblock
\shownote{Accessed: September 25, 2024}.


\bibitem[Tau~Leng et~al\mbox{.}(2002)]%
        {tau2002empirical}
\bibfield{author}{\bibinfo{person}{Rizwan~Ali Tau~Leng}, \bibinfo{person}{Jenwei Hsieh}, \bibinfo{person}{Victor Mashayekhi}, {and} \bibinfo{person}{Reza Rooholamini}.} \bibinfo{year}{2002}\natexlab{}.
\newblock \showarticletitle{An empirical study of hyper-threading in high performance computing clusters}.
\newblock \bibinfo{journal}{\emph{Linux HPC Revolution}}  \bibinfo{volume}{45} (\bibinfo{year}{2002}).
\newblock


\bibitem[Thom et~al\mbox{.}(2018)]%
        {thom2018survey}
\bibfield{author}{\bibinfo{person}{Mark Thom}, \bibinfo{person}{Gerhard~W Dueck}, \bibinfo{person}{Kenneth Kent}, {and} \bibinfo{person}{Daryl Maier}.} \bibinfo{year}{2018}\natexlab{}.
\newblock \showarticletitle{A survey of ahead-of-time technologies in dynamic language environments}. In \bibinfo{booktitle}{\emph{Proceedings of the 28th Annual International Conference on Computer Science and Software Engineering}}. \bibinfo{pages}{275--281}.
\newblock


\bibitem[Tolley(2022)]%
        {pyjion}
\bibfield{author}{\bibinfo{person}{Anthony Tolley}.} \bibinfo{year}{2022}\natexlab{}.
\newblock \bibinfo{title}{{Pyjion}: A {JIT} compiler for {CPython}}.
\newblock \bibinfo{howpublished}{\url{https://github.com/tonybaloney/pyjion}}.
\newblock
\newblock
\shownote{Accessed: 2024-01-14}.


\bibitem[van Kempen et~al\mbox{.}(2024)]%
        {van2024s}
\bibfield{author}{\bibinfo{person}{Nicolas van Kempen}, \bibinfo{person}{Hyuk-Je Kwon}, \bibinfo{person}{Dung~Tuan Nguyen}, {and} \bibinfo{person}{Emery~D Berger}.} \bibinfo{year}{2024}\natexlab{}.
\newblock \showarticletitle{It's Not Easy Being Green: On the Energy Efficiency of Programming Languages}.
\newblock \bibinfo{journal}{\emph{arXiv preprint arXiv:2410.05460}} (\bibinfo{year}{2024}).
\newblock


\bibitem[Weber et~al\mbox{.}(2023)]%
        {weber2023twins}
\bibfield{author}{\bibinfo{person}{Max Weber}, \bibinfo{person}{Christian Kaltenecker}, \bibinfo{person}{Florian Sattler}, \bibinfo{person}{Sven Apel}, {and} \bibinfo{person}{Norbert Siegmund}.} \bibinfo{year}{2023}\natexlab{}.
\newblock \showarticletitle{Twins or false friends? a study on energy consumption and performance of configurable software}. In \bibinfo{booktitle}{\emph{2023 IEEE/ACM 45th International Conference on Software Engineering (ICSE)}}. IEEE, \bibinfo{pages}{2098--2110}.
\newblock


\bibitem[Wohlin et~al\mbox{.}({[n.\,d.]})]%
        {wohlinexperimentation}
\bibfield{author}{\bibinfo{person}{Claes Wohlin}, \bibinfo{person}{Per Runeson}, \bibinfo{person}{Martin H{\"o}st}, \bibinfo{person}{Magnus~C Ohlsson}, \bibinfo{person}{Bj{\"o}rn Regnell}, \bibinfo{person}{Anders Wessl{\'e}n}, {et~al\mbox{.}}} \bibinfo{year}{[n.\,d.]}\natexlab{}.
\newblock \showarticletitle{Experimentation in Software Engineering [electronic resource]: An Introduction}.
\newblock  (\bibinfo{year}{[n.\,d.]}).
\newblock


\bibitem[Zhang et~al\mbox{.}(2022a)]%
        {zhang2022regcpython}
\bibfield{author}{\bibinfo{person}{Qiang Zhang}, \bibinfo{person}{Lei Xu}, {and} \bibinfo{person}{Baowen Xu}.} \bibinfo{year}{2022}\natexlab{a}.
\newblock \showarticletitle{RegCPython: A Register-based Python Interpreter for Better Performance}.
\newblock \bibinfo{journal}{\emph{ACM Transactions on Architecture and Code Optimization}} \bibinfo{volume}{20}, \bibinfo{number}{1} (\bibinfo{year}{2022}), \bibinfo{pages}{1--25}.
\newblock


\bibitem[Zhang et~al\mbox{.}(2024)]%
        {zhang2024python}
\bibfield{author}{\bibinfo{person}{Qiang Zhang}, \bibinfo{person}{Lei Xu}, {and} \bibinfo{person}{Baowen Xu}.} \bibinfo{year}{2024}\natexlab{}.
\newblock \showarticletitle{Python meets JIT compilers: A simple implementation and a comparative evaluation}.
\newblock \bibinfo{journal}{\emph{Software: Practice and Experience}} \bibinfo{volume}{54}, \bibinfo{number}{2} (\bibinfo{year}{2024}), \bibinfo{pages}{225--256}.
\newblock


\bibitem[Zhang et~al\mbox{.}(2022b)]%
        {zhang2022quantifying}
\bibfield{author}{\bibinfo{person}{Qiang Zhang}, \bibinfo{person}{Lei Xu}, \bibinfo{person}{Xiangyu Zhang}, {and} \bibinfo{person}{Baowen Xu}.} \bibinfo{year}{2022}\natexlab{b}.
\newblock \showarticletitle{Quantifying the interpretation overhead of Python}.
\newblock \bibinfo{journal}{\emph{Science of Computer Programming}}  \bibinfo{volume}{215} (\bibinfo{year}{2022}), \bibinfo{pages}{102759}.
\newblock


\end{thebibliography}
\end{document}